\documentclass[pra,twocolumn,preprintnumbers,amsmath,amssymb,superscriptaddress]{revtex4-2}
\usepackage{dsfont}
\usepackage{color}
\usepackage{graphicx}
\usepackage{dcolumn}
\usepackage{bm}
\usepackage{hyperref}
\usepackage{enumitem}
\usepackage{textcomp}
\usepackage{balance}
\usepackage{comment}
\usepackage{cleveref}
\usepackage{bbold}
\usepackage{amsthm}

\begin{document}

\title{Macroscopic quantum teleportation with ensembles of qubits}

\author{Manish Chaudhary}
\thanks{The indicated authors are joint first authors}
\affiliation{Institut de Physique Nucléaire, Atomique et de Spectroscopie, CESAM, University of Liège, B-4000 Liège, Belgium}
\affiliation{State Key Laboratory of Precision Spectroscopy, School of Physical and Material Sciences, East China Normal University, Shanghai 200062, China}
\affiliation{New York University Shanghai; NYU-ECNU Institute of Physics at NYU Shanghai, 567 West Yangsi Road, Pudong, Shanghai 200126, China}

\author{Zhiyuan Lin} 
\thanks{The indicated authors are joint first authors}
\affiliation{State Key Laboratory of Precision Spectroscopy, School of Physical and Material Sciences, East China Normal University, Shanghai 200062, China}
\affiliation{New York University Shanghai; NYU-ECNU Institute of Physics at NYU Shanghai, 567 West Yangsi Road, Pudong, Shanghai 200126, China}

\author{Shuang Li} 
\affiliation{State Key Laboratory of Precision Spectroscopy, School of Physical and Material Sciences, East China Normal University, Shanghai 200062, China}
\affiliation{New York University Shanghai; NYU-ECNU Institute of Physics at NYU Shanghai, 567 West Yangsi Road, Pudong, Shanghai 200126, China}

\author{Mohan Zhang} 
\affiliation{State Key Laboratory of Precision Spectroscopy, School of Physical and Material Sciences, East China Normal University, Shanghai 200062, China}
\affiliation{New York University Shanghai; NYU-ECNU Institute of Physics at NYU Shanghai, 567 West Yangsi Road, Pudong, Shanghai 200126, China}

\author{Yuping Mao}
\affiliation{Department of Physics, Huzhou Teachers' College, Huzhou 313000, China}
\affiliation{State Key Laboratory of Precision Spectroscopy, School of Physical and Material Sciences, East China Normal University, Shanghai 200062, China}
\affiliation{New York University Shanghai; NYU-ECNU Institute of Physics at NYU Shanghai, 567 West Yangsi Road, Pudong, Shanghai 200126, China}

\author{Valentin Ivannikov} 
\affiliation{New York University Shanghai; NYU-ECNU Institute of Physics at NYU Shanghai, 567 West Yangsi Road, Pudong, Shanghai 200126, China}

\author{Tim Byrnes}
\email{tim.byrnes@nyu.edu}
\affiliation{New York University Shanghai; NYU-ECNU Institute of Physics at NYU Shanghai, 567 West Yangsi Road, Pudong, Shanghai 200126, China}
\affiliation{State Key Laboratory of Precision Spectroscopy, School of Physical and Material Sciences, East China Normal University, Shanghai 200062, China}
\affiliation{Center for Quantum and Topological Systems (CQTS),
NYUAD Research Institute, New York University Abu Dhabi, UAE}
\affiliation{Department of Physics, New York University, New York, NY, 10003, USA}

\date{\today}

\begin{abstract}
We develop methods for performing quantum teleportation of the total spin variables of an unknown state, using quantum nondemolition measurements, spin projection measurements, and classical communication.  While theoretically teleportation of high-dimensional states can be attained with the assumption of generalized Bell measurements, this is typically experimentally non-trivial to implement.  We introduce two protocols and show that, on average, the teleportation succeeds in teleporting the spin variables of a spin coherent state with average zero angular error in the ideal case, beating classical strategies based on quantum state estimation.  In a single run of the teleportation, there is an angular error at the level of $ \sim 0.1 $ radians for large ensembles.  A potential physical implementation for the scheme is with atomic ensembles and quantum nondemolition measurements performed with light. We analyze the decoherence of the protocols and find that the protocol is robust even in the limit of large ensemble sizes.  
\end{abstract}

\maketitle

\section{\label{sec1}Introduction}

One of the most spectacular applications of entanglement is quantum teleportation \cite{bennett1993teleporting,pirandola2015advances}.  
In quantum teleportation, an unknown quantum state is transmitted to a  distant party using entanglement and classical communication \cite{pirandola2015advances,hu2023progress}. 
The protocol for quantum teleportation has been demonstrated in numerous pioneering experiments \cite{bouwmeester1997experimental,furusawa1998unconditional} with optical \cite{boschi1998experimental,PhysRevLett.86.1370,jin2010experimental}, trapped ions \cite{riebe2004deterministic,barrett2004deterministic,olmschenk2009quantum,daiss2021quantum,wan2019quantum}, superconducting qubits \cite{steffen2013deterministic} and atomic 
\cite{bao2012quantum,krauter2013deterministic} systems.  
Significant progress has been made in establishing long-distance quantum communication using quantum teleportation based on near-Earth satellites \cite{yin2012quantum,ma2012quantum,ren2017ground}.   It is also an essential step in numerous applications such as distributed quantum computing \cite{PhysRevA.76.062323,PhysRevA.89.022317,caleffi2024distributed,cacciapuoti2019quantum}, and operations such as gate teleportation \cite{gottesman1999demonstrating,wan2019quantum} which form the foundation of measurement-based quantum computing \cite{raussendorf2001one}.   

The simplest version of the teleportation protocol involves teleporting a two-level system, i.e. a qubit.  Extending this to higher dimensional systems has always been a topic of fascination among physicists, but has thus far remained experimentally challenging.  The largest discrete system that has been teleported, at the time of writing is a qutrit (3-level) photonic system \cite{luo2019quantum,hu2020experimental}.  A two qubit system was teleported with photons in Ref. \cite{zhang2006experimental,PhysRevA.82.032318,PhysRevA.92.042314}.  Teleportation of encoded state has been achieved in works such as Ref. \cite{huang2021emulating,ryan2024high,zhang2022resource} where a qubit was encoded in higher dimensional system. Theoretically, performing high-dimensional teleportation follows the same procedure as the qubit case \cite{bennett1993teleporting}.  First, a pair of high-dimensional entangled systems is prepared, then a maximally entangled measurement in the high dimensional space is performed. Conditional classical unitary operations then complete the teleportation process.  The problem with this is that generating the high-dimensional entangled state and/or realizing the entangled basis measurement are not easily performed.  Furthermore, there is the risk that high-dimensional systems suffer decoherence at an accelerated rate in comparison to qubits, such that any quantum effects are removed \cite{zurek2003decoherence}. For these reasons, performing high-dimensional teleportation remains a challenging problem.


Atomic ensembles constitute a natural system that is suitable for realizing quantum operations on high-dimensional states, due to the precise controllability and low decoherence rates \cite{hammerer2010quantum,lukin2000entanglement}. One of the limitations of such systems is the lack of individual control of the underlying atoms, hence the only operations that are available are global operations applied to the entire ensemble.  For this reason, these have been mainly considered for applications such as quantum metrology \cite{gross2012spin}.  Most of the work relating to entanglement has been focused on a single ensemble, where a large focus has been to generate squeezed states for measurements exceeding the standard quantum limit \cite{kitagawa1993squeezed,hald1999spin,sewell2012magnetic}. In Ref. \cite{schmied2016bell} Bell correlations within a single Bose-Einstein condensate (BEC) were found, which was then subsequently extended to two spatial regions of the same BEC \cite{fadel2018spatial,kunkel2018spatially,lange2018entanglement}.  Generating entanglement between two spatially separate atomic ensembles was performed in Ref. \cite{julsgaard2001experimental} and more recently also in Bose-Einstein condensates \cite{colciaghi2023einstein}.  Entanglement between four ensembles was recently performed in Ref. \cite{cooper2024graph}.  Such entanglement between two or more ensembles is interesting from the perspective of generalizing spin squeezing to a multi-ensemble context \cite{byrnes2013fractality,kitzinger2020,byrnes2024multipartite}, and also performing various quantum information tasks based on atomic ensembles, such as multiparameter quantum metrology \cite{fadel2022multiparameter}, quantum teleportation \cite{krauter2013deterministic,pyrkov2014quantum,pyrkov2014full,braunstein2005quantum}, remote state preparation \cite{manish2021remote}, clock synchronization \cite{ilo2018remote}, and quantum computing \cite{byrnes2012macroscopic,byrnes2015macroscopic,abdelrahman2014coherent}.

\begin{figure}%
\includegraphics[width=\linewidth]{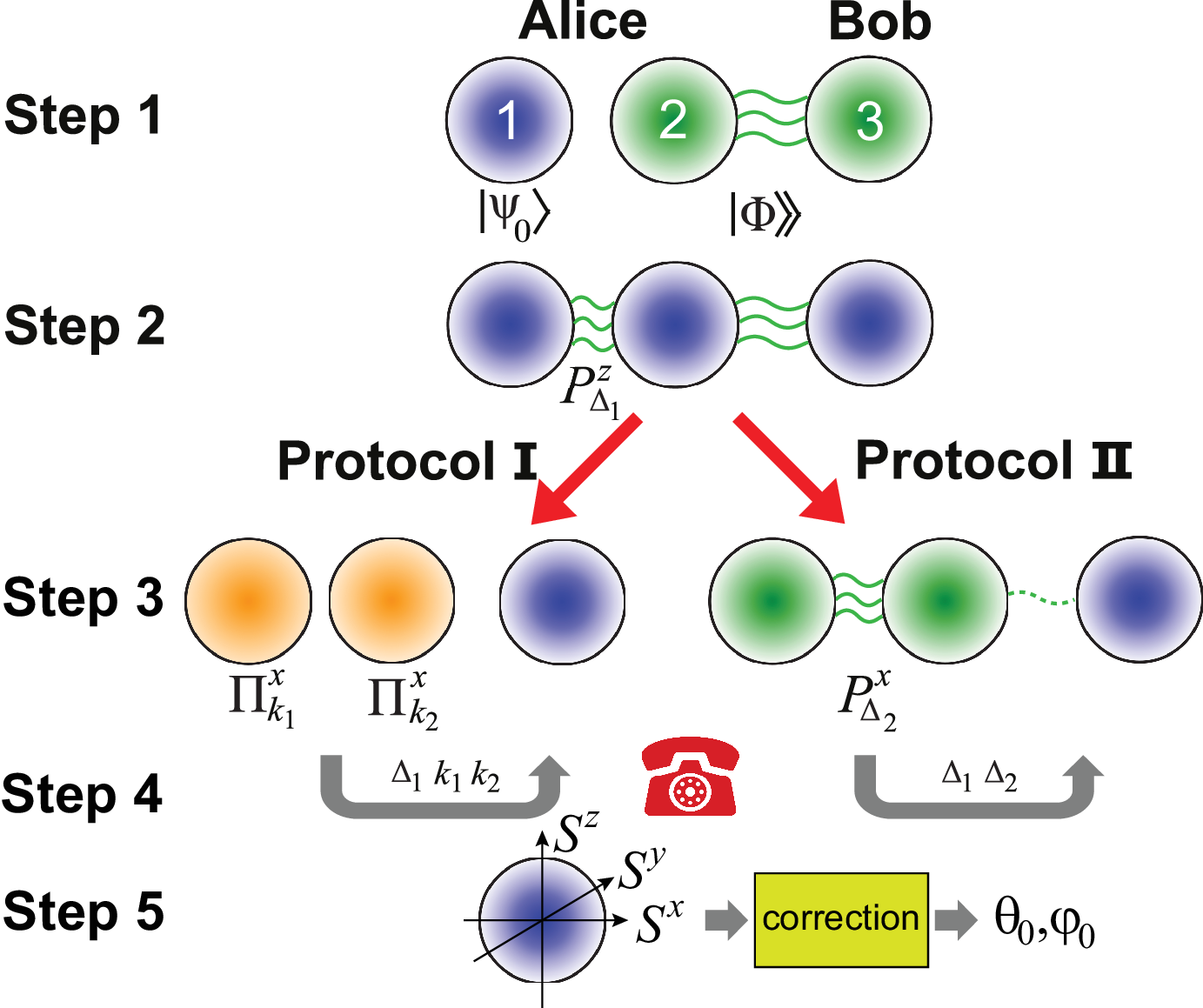}
	\caption{Teleportation protocols for transmitting the spin variables of an initial spin coherent state from Alice to Bob.  Alice is posession of ensembles 1 and 2, while Bob has ensemble 3.  The steps involved in Protocol I and II, detailed in Sec. \ref{sec3}, are summarized pictorially. 
In step 1, the initial state  $ | \theta_0, \phi_0 \rangle \rangle   \otimes | \Phi \rangle \rangle $ is prepared.  In step 2, a QND projective measurement in the $z$-basis is performed.  At this point the 3 ensembles are entangled (indicated by the wavy lines).  For Protocol I, in step 3 Alice's ensembles are measured individually in the $ x $ basis.  In Protocol II, Alice's ensembles are measured by a second QND projective measurement in the $ x $-basis.  In the case of Protocol I, ensemble 3 disentangles from the other ensembles.  In the case of Protocol II, there is some remnant entanglement with ensemble 3.  In step 4, the outcomes of the measurements are transmitted classically from Alice to Bob.  In Step 5, the spins of the output state are measured and processed classically to recover the initial spin coherent state parameters $ \theta_0, \phi_0  $.  }
	\label{fig1}%
\end{figure}

In this paper, we introduce two protocols to teleport the spin variables of an unknown state on a qubit ensemble. 
In constructing our protocol, we restrict ourselves to only experimentally viable operations, namely QND measurements \cite{aristizabal2021quantum,manish2022measurement},  spin projection measurements, and classical communications.  Our protocols are illustrated in Fig. \ref{fig1}.  As with standard qubit teleportation, Alice is in possession of two ensembles, one containing the original states to be teleported and the other a maximally entangled state that is shared with Bob. To generate the maximally entangled state, we use the protocol introduced in our earlier paper Ref. \cite{manish2023maximallyentangled}, where the entangled state between two ensembles can be prepared using only QND measurements and local spin rotations. The two variants of the protocols (called I and II), give similar results, but differ in their details and the correction protocol. In Protocol I, one QND measurement is performed, followed by local spin measurements on the two ensembles.  In Protocol II, two QND measurements are made, in the $ z $, then in the $ x $ basis.  Finally, a spin readout of Bob's ensemble is made, and combined with correction operations based on the measurement outcomes, this completes the teleportation procedure. 
We find that on average, the spin readout gives remarkably {\it perfect} transfer of Alice's original spin variables for a spin coherent state. We limit the teleportation to spin variables, as opposed to the full quantum state, due to the restricted operations that are available in the context of qubit ensembles.  We however note that the task of teleporting spin averages is important in the context of spinor quantum computing 
\cite{byrnes2012macroscopic,timquantumoptics2020}, where spin coherent states are used to encode qubits.

We make a brief contrast of our teleportation protocols to past works.  
In the context of atomic ensembles, there have been two approaches to perform teleportation, one being continuous variable teleportation \cite{krauter2013deterministic,sherson2006quantum} and the other using collective excitations \cite{bao2012quantum}.  For continuous variables teleportation, the approach is to use the Holstein-Primakoff (HP) approximation to effectively work within the phase space of position and momentum \cite{braunstein2005quantum}. This works in a small region of Hilbert space, examining small deviations from a polarized spin direction. For example, working with an ensemble polarized in the $ S^x $ direction, under the HP approximation one is constrained to Bloch sphere angles in the region of $ \theta = \pi/2 \pm O(1/\sqrt{N})$ and $ \phi = 0 \pm O(1/\sqrt{N}) $. In contrast, our teleportation is designed to work for arbitrary $ \theta, \phi $. Collective excitations also work in a similar way, by polarizing the spins in an ensemble, then looking at discrete excitations of this.  Our protocol differs from these scheme in that the aim is to teleport states with an arbitrary polarization on the Bloch sphere.  Previously, we proposed two ``beyond-HP'' schemes, for teleportation of the equatorial \cite{pyrkov2014quantum} and full Bloch sphere coordinates \cite{pyrkov2014full}.  These were both however based on $ S^z S^z $ interactions between ensembles, which are still yet to be experimentally realized for remote ensembles in a controllable way.  Therefore, to our knowledge, there is still no practically realizable scheme for beyond-HP teleportation.

The paper is structured as follows.  In Sec. \ref{sec2} we review the 
physical system and the operations that can be carried out, including a summary of the protocol for preparing a maximally entangled state of two ensembles \cite{manish2023maximallyentangled}.  In Sec. \ref{sec3} we introduce the protocols for quantum teleportation for macroscopic atomic ensembles.  These are analyzed mathematically in Sec. \ref{sec4}, with the associated justification for the protocols.   In Sec. \ref{sec5} we numerically simulate our proposed protocols and show the overall performance of the teleportation. In Sec. \ref{sec6} we compare it with the classical bounds for transmitting the quantum information and show that in the ideal case our protocol outperforms them.  In Sec. \ref{sec7} we analyze the effect of decoherence on the efficiency of the teleportation protocols. In Sec. \ref{sec8} we briefly discuss the experimental implementation in the context of atomic ensembles.   Finally, in Sec. \ref{sec9} we give a summary and provide the main conclusions of our study.

\section{Physical system and required operations}
\label{sec2}

Here we introduce the necessary formalism to capture the physical system in question and the required operations to carry out the quantum teleportation protocol.

\subsection{Qubit ensembles}

The physical system we consider are ensembles consisting of $N$ non-interacting qubits. Quantum information is stored on the collective spin operators in terms of the sum of individual spin operators 
\begin{align}
    S^\eta_l & =  \sum_{n=1}^N \sigma^x_{l,n} 
     \label{eq:spinopsensemble}
\end{align}
where $ \eta \in \{ x,y,z \} $ and $ \sigma^\eta_{l,n} $ is a Pauli operator in the $ \eta $ direction for the $n^{\text{th}}$ atom in the $l^{\text{th}}$ ensemble. We consider the number of atoms $N$ in each ensemble to be equal for simplicity. To keep the discussion as general as possible, we do not further specify the physical system, beyond that the decoherence to be at a sufficiently low level such that the quantum operations in the protocol can be performed coherently.  Qubit ensembles can be physically implemented by neutral atomic gas ensembles, for example.  More details on the experimental implementation is deferred to Sec. \ref{sec7}.  The commutation relation for the spin operators are
\begin{align}
[S^i_l,S^j_m]=2i\epsilon_{ijk} S^k \delta_{lm},   \label{commutator}
\end{align}
where $\epsilon_{ijk}$ is the Levi-Civita symbol.

The spin operators (\ref{eq:spinopsensemble}) are symmetric under particle interchange.  This is a important limitation of working with such ensembles, that the underlying particles are not addressable, but each ensemble is individually controllable.  It then follows that if the initial state is symmetric under particle interchange and only symmetric operators are used to manipulate the state, the state remains in the symmetric subspace \cite{timquantumoptics2020}. We may then equivalently work using Schwinger boson operators 
\begin{align}\label{eq:spinops}
    S^x_l & =a_l^\dagger b_l+b_l^\dagger a_l \nonumber \\
    S^y_l & =-ia_l^\dagger b_l+ib_l^\dagger a_l \nonumber \\
     S^z_l & =a_l^\dagger a_l-b_l^\dagger b_l  . 
\end{align}
where the bosonic annihilation operator for the qubit states are $a_l, b_l $ respectively.  Working in the completely symmetric subspace reduces the Hilbert space dimension of each ensemble from $ 2^N $ to $ N +1 $.  For convenience we shall use the bosonic formulation (\ref{eq:spinops}) henceforth.  

A spin coherent state for $N$ uncorrelated qubits in an ensemble is defined as
\begin{align}
    | \theta , \phi \rangle\rangle_l =\frac{( \cos \frac{\theta}{2} e^{-i \phi /2 }  a_l^\dagger+ \sin \frac{\theta}{2} e^{i \phi /2 } b_l^\dagger)^N}{\sqrt{N!}}| \text{vac} \rangle
    \label{spincoherent}
\end{align}
where $\theta , \phi$ are the angles on the Bloch sphere, and $ | \text{vac} \rangle $ is the vacuum state containing no atoms. The subscript $ l $ on states refers to the ensemble number throughout this paper.  

The number (or Dicke) states are defined as 
\begin{align}
|k \rangle_l = \frac{(a_l^\dagger)^k (b_l^\dagger)^{N-k}}{\sqrt{k! (N-k)!}} | \text{vac} \rangle .
\label{dickestates}
\end{align}
The number states are eigenstates of the $ S^z $ operator defined as
\begin{align}
S^z_l |k \rangle_l =  (2 k-N) |k \rangle_l  .
\end{align}

\subsection{Ensemble operations}
\label{sec:qnd}

Here we list the operations that we assume are available on the qubit ensembles.  

We assume that each of the ensembles can be manipulated individually with total spin rotations
\begin{align}
U_l(\vec{n} ,\varphi) = e^{-i \vec{n} \cdot \vec{S}_l \varphi/2 }
\label{oneensrot}
\end{align}
where $ \vec{n} $ is a three dimensional unit vector that specifies the axis of the rotation, $  \vec{S}_l = ( S^x_l,S^y_l,S^z_l) $, and $ \varphi $ is the angle of rotation.  

In addition, measurements in the number basis are assumed to be possible, corresponding to the measurement operator
\begin{align}
\Pi_{k}^{z,l} = |k \rangle_l \langle k |_l ,
\label{projz}
\end{align}
where the outcome is $ k $ and the $ z $ denotes a measurement in the $ S^z $ eigenbasis. The superscript $ l $ indicates which ensemble the projector acts on. Measurements in the $ x $ basis correspond to 
\begin{align}
\Pi_{k}^{x,l} = |k \rangle_l^{(x)} \langle k |_l^{(x)}.
\label{projx}
\end{align}
where the number states in the $ x $ basis are defined as 
\begin{align}
  |k \rangle^{(x)}_l =  e^{-i S^y_l \pi/4} |k \rangle_l  .
  \label{kxbasisdef}
\end{align}
These are projection operators and satisfy
\begin{align}
\sum_{k=0}^N \Pi_{k}^{z,l}  = \sum_{k=0}^N \Pi_{k}^{x,l} = I_{1}  
\end{align}
where $I_{1} $ is the identity matrix for a single ensemble with dimension $ N+1 $. 

To entangle the ensembles, we assume that QND measurements can be performed on two ensembles. Under particular conditions (see  Appendix \ref{app:qnd}), this can be written as a projection operator
\begin{align}
P_\Delta^{z} =  \sum_{k=\max(0,-\Delta)}^{\min(N,N-\Delta) }  |k,k + \Delta\rangle \langle k,k+\Delta|  , 
\label{projzbasis_singlepeak}
\end{align}
where the $ z $ refers to the basis that the number states are in. 

We may also perform QND measurements in the $ x $ basis according to
\begin{align}
  P_\Delta^{x}=  e^{- i (S^x_1+ S^x_2) \pi/4 } P_\Delta^{z}  e^{i (S^x_1+ S^x_2) \pi/4 } ,
\label{projzbasis_singlepeak_x}
\end{align}
where the exponential factors are single ensemble rotations (\ref{oneensrot}). Through this paper, the QND measurements always act on ensembles 1 and 2, hence we omit the ensemble labels on the left hand side of (\ref{projzbasis_singlepeak_x}).  The projection operators satisfy
\begin{align}
    \sum_{\Delta=-N}^N P_\Delta^{z} = \sum_{\Delta=-N}^N P_\Delta^{x} = I_{2} ,
\end{align}
where $ I_2 $ is the identity matrix of two ensembles with dimension $ (N+1)^2 $.

\subsection{Maximally entangled state preparation}
\label{sec:mmes}

One of the key resources that is required in the quantum teleportation is shared entanglement between Alice and Bob.  For this we prepare a maximally entangled state between two ensembles shared by Alice and Bob (see Fig. \ref{fig1}). We define the maximally entangled state as  \cite{kitzinger2020,byrnes2024multipartite}
\begin{align}
|\Phi \rangle\rangle & = \frac{1}{N! \sqrt{N+1}} (a_1^\dagger a_2^\dagger + b_1^\dagger b_2^\dagger )^N | \text{vac} \rangle \\
& = \frac{1}{\sqrt{N+1}} \sum_{k=0}^N |k\rangle \otimes |k \rangle . 
\label{mmesstate}
\end{align}
In Ref. \cite{manish2023maximallyentangled}, a deterministic method of preparing this state was introduced.  The method uses only the QND measurements (\ref{projzbasis_singlepeak}), (\ref{projzbasis_singlepeak_x}), and conditional unitary rotations (\ref{oneensrot}).  We summarize how this is performed in Appendix \ref{app:maxent}.

\section{Teleportation protocols}
\label{sec3}

In this section we introduce two teleportation protocols (called Protocol I and II) for transmitting the total spin information of an unknown ensemble. Here, we simply give a concrete description of the protocols.  The theoretical analysis, and the rationale of the steps will be performed in Sec. \ref{sec4}.

\subsection{Protocol I}
\label{sec:protocolI}

Protocol I involves performing one QND measurement in the $ z $-basis, followed by local number measurements of Alice's ensembles in the $ x $-basis (see Fig. \ref{fig1}).  The specific steps are as follows.
\begin{enumerate}
   \item Alice prepares her initial state on ensemble 1 and shares a maximally entangled state with Bob:
   \begin{align}
    |\Psi_0\rangle & = | \psi_0 \rangle_1 \otimes | \Phi \rangle \rangle_{23}  .
    \label{initialstate}
 \end{align}

\item A QND measurement in the $z$-basis is performed on ensembles 1 and 2, which yields the state $ P_\Delta^{z} | \Psi_0\rangle  $. 

\item A local measurement in the number state basis is made on ensembles 1 and 2.  This produces the state $ \Pi_{k_1}^{x,1} \Pi_{k_2}^{x,2} P_\Delta^{z} |\Psi_0\rangle  $. 

\item Alice transmits the outcomes of the measurement $ (\Delta, k_1, k_2 )$ to Bob. 

\item Bob measures the expectation value of $ \vec{S}_3 $ and performs a correction according to
\begin{align}
S^x_{\text{tel},I} &  = (-1)^{ H(k_1 - \frac{N}{2} ) + H(k_2 - \frac{N}{2} ) } \langle S^x_3 \rangle \nonumber \\
S^y_{\text{tel},I} &  = (-1)^{ H(k_1 - \frac{N}{2} ) + H(k_2 - \frac{N}{2} ) } \langle S^y_3 \rangle  \nonumber  \\
S^z_{\text{tel},I} &  = \langle S^z_3 \rangle - 2 \Delta ,
\label{spintelprot1}
\end{align}
where $ H(x) $ is the Heaviside step function with $ H(x) = 1$ for $ x\ge 0 $ and $ H(x) = 0$ otherwise.  The Bloch sphere angles can be found by
\begin{align}
\theta_{\text{tel},I} = &  \cos^{-1} \left( \frac{ S^z_{\text{tel},I} }{N} \right) \label{thetatelpro1} \\
\phi_{\text{tel},I} = & \tan^{-1} \left( 
\frac{S^y_{\text{tel},I}  }{S^x_{\text{tel},I}}  \right) \nonumber \\
& + \pi ( H(k_1 - \frac{N}{2} ) + H(k_2 - \frac{N}{2} ) ) . \label{phitelpro1} 
\end{align}
\end{enumerate}

%

\subsection{Protocol II}

Protocol II involves performing two QND measurements in the $ z $-basis and $ x $-basis in that order (see Fig. \ref{fig1}).  The specific steps are as follows.  
\begin{enumerate}
   \item Alice prepares her initial state on ensemble 1 and shares a maximally entangled state with Bob:
   \begin{align}
    |\Psi_0\rangle & = | \psi_0 \rangle_1 \otimes | \Phi \rangle \rangle_{23}  .
 \end{align}

\item A QND measurement in the $z$-basis is performed on ensembles 1 and 2, which yields the state $ P_{\Delta_1}^{z} | \Psi_0\rangle  $. 

\item A QND measurement in the $x$-basis is performed on ensembles 1 and 2, which yields the state $ P_{\Delta_2}^{x} P_{\Delta_1}^{z} | \Psi_0\rangle  $. 

\item Alice transmits the outcomes of the measurement $ (\Delta_1, \Delta_2 )$ to Bob. 

\item Bob measures the expectation value of $ \vec{S}_3 $ and performs a correction according to
\begin{align}
S^x_{\text{tel},II} &  = (-1)^{ H( |\Delta_2 | - \Delta_{+}  )} \langle S^x_3 \rangle \nonumber \\
S^y_{\text{tel},II} &  = (-1)^{ H( |\Delta_2 | - \Delta_{+}  )} \langle S^y_3 \rangle \nonumber \\
S^z_{\text{tel},II} &  = \langle S^z_3 \rangle - 2 \Delta ,
\label{spintelprot2}
\end{align}
where
\begin{align}
    \Delta_{+} = \frac{1}{2} \left( -N -1 + \sqrt{3N^2 + 6N + 1} \right) .
\end{align}
The Bloch sphere angles can be found by
\begin{align}
\theta_{\text{tel},II} = &  \cos^{-1} \left( \frac{ S^z_{\text{tel},II} }{N} \right) \label{thetatelpro2} \\
\phi_{\text{tel},II} = & \tan^{-1} \left( 
\frac{S^y_{\text{tel},II}  }{S^x_{\text{tel},II}} \right) +  \pi H \Big( |\Delta_2 | -  \Delta_{+} \Big) .  \label{phitel2}
\end{align}
\end{enumerate}

We point out that the correction operations that are performed in (\ref{spintelprot1}) and (\ref{spintelprot2}) both only require the measurement outcomes $ k_1, k_2, \Delta $ for Protocol I and $\Delta_1, \Delta_2 $ for Protocol II.  As such, the protocols  can be executed without Alice or  Bob's knowledge of the state $ | \Psi_0 \rangle $.

\section{Analysis of the teleportation protocols}
\label{sec4}

In this section we analyze the quantum states generated by the two teleportation protocols introduced in Sec. \ref{sec3} and give an explanation of the origin of Bob's correction operations.

\begin{figure}%
\includegraphics[width=\linewidth]{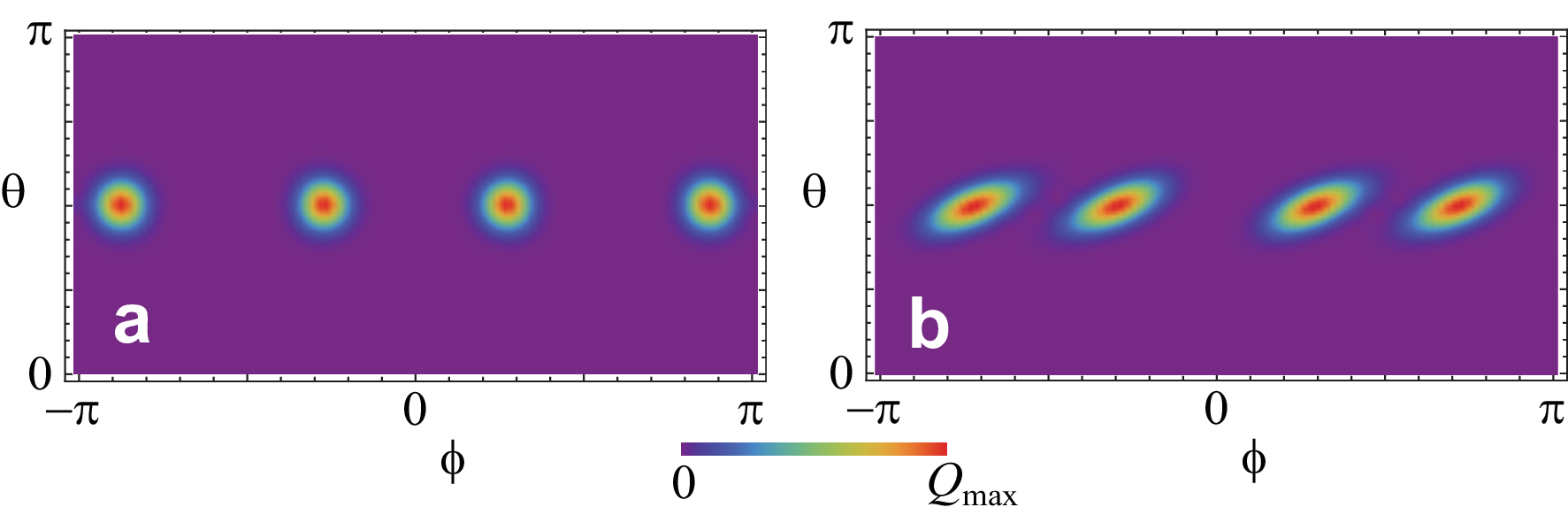}
	\caption{Q functions (\ref{qfuncprotocol1}) for the quantum state $ | \Psi_I \rangle $ given by (\ref{step3protocol1}).  The initial states $ | \psi_0 \rangle $ are (a) $ | \theta = \pi/2, \phi = 0 \rangle \rangle $ and (b) $ e^{i (S^z)^2 /(2N) } | \theta = \pi/2, \phi = 0 \rangle \rangle $.  The measurement outcomes are (a) $ k_1 = 20, k_2 = 60 $, (b) $ k_1 = 10, k_2 = 50$.  We take $ N = 100 $ for all the graphs.  }
	\label{fig2}%
\end{figure}

\subsection{Protocol I}

We now analyze Protocol I by giving the expressions for the quantum state.

The unnormalized state after step 3 of the protocol evaluates to
\begin{align}
| \Psi_I \rangle =  \Pi_{k_1}^{x,1} \Pi_{k_2}^{x,2}  P_{\Delta}^{z} |\Psi_0\rangle  = \frac{1}{\sqrt{N+1}} |k_1 \rangle^{(x)}_1 |k_2 \rangle^{(x)}_2 \nonumber \\
  \otimes \sum_{k=\max(0,-\Delta)}^{\min(N,N-\Delta) } 
\psi_k^{(0)} R_{k_1,k} R_{k_2,k+\Delta}  |k+\Delta \rangle_{3}  ,
\label{step3protocol1}
\end{align}
where $ \psi_k^{(0)}  = \langle k | \psi_0 \rangle $ and we defined the matrix elements
\begin{align}
R_{k',k} = \langle k' |^{(x)}  | k \rangle .
\end{align}
The matrix elements can be evaluated using the expression in Appendix \ref{app:prob}.
After the $ x $-basis projections, we can see that the ensemble 3 is disentangled from the other two ensembles.  

The probability of obtaining this measurement outcome is
\begin{align}
p_I (k_1, k_2, \Delta) & = \langle \Psi_I | \Psi_I \rangle \nonumber \\
& = \frac{1}{N+1}  \sum_{k=\max(0,-\Delta)}^{\min(N,N-\Delta) } 
|\psi_k^{(0)}R_{k_1,k} R_{k_2,k+\Delta} |^2 
\label{probprot1}
\end{align}

The factors of $ R $ and the offset $ \Delta $ of the number state in (\ref{step3protocol1}) give a modification of the initial state $ | \psi_0 \rangle $ originally on ensemble 1.  To better understand the nature of these corrections, we make an approximation for the $ R $ matrix elements using the expression (\ref{finalapproxOmega}) 
\begin{align}
R_{k',k}\approx  A_{k',k}  ( e^{i \phi_{k'} (2 k - N)/2 } + 
(-1)^{n_{k'}} e^{-i \phi_{k'} (2 k - N) /2}  )
\label{rapprox}
\end{align}
where we defined
\begin{align}
\phi_{k'} = \cos^{-1} \Big ( \frac{2k'}{N} -1 \Big) .
\label{phidef}
\end{align}
and
\begin{align}
n_k = \frac{N}{2} - \left| k - \frac{N}{2} \right| .
\label{nkdef}
\end{align}
and the function $ A_{k', k} $ is defined in (\ref{amplitudefunc}). 

Substituting the approximation (\ref{rapprox}) into (\ref{step3protocol1}), we obtain
\begin{align}
& | \Psi_I \rangle \approx  \sum_{k=\max(0,-\Delta)}^{\min(N,N-\Delta) } A_{k_1,k} A_{k_2,k+\Delta} 
\psi_k^{(0)}  \nonumber \\
& \times ( e^{i \phi_{k_1} (2 k - N)/2 } + 
(-1)^{n_{k_1}} e^{-i \phi_{k_1} (2 k - N) /2}  ) \nonumber  \\
& \times ( e^{i \phi_{k_2} (2 k + 2 \Delta - N)/2 } + 
(-1)^{n_{k_2}} e^{-i \phi_{k_2} (2 k + 2 \Delta - N) /2}  ) |k+\Delta \rangle_{3}  ,
\end{align}
where we have dropped the states on ensembles 1 and 2 and overall constant factors which do not affect the quantum state on ensemble 3.  

The amplitude function $  A_{k',k} $ is a slowly varying function of $ k $ which only affects the wavefunction if $ \psi_k^{(0)} $ is broad in the $ k $ variable.  For teleporting states such as spin coherent states, $ \psi_k^{(0)} $ is strongly peaked which makes the dominant effect in the phase factors in (\ref{rapprox}). Ignoring these factors and only retain the phase factors we obtain
\begin{align}
 | \Psi_I \rangle & \approx  \frac{1}{2} 
\Big[ e^{-i \phi_{k_1} \Delta}  e^{i (\phi_{k_1} + \phi_{k_2} )S^z/2 } \nonumber \\
& + (-1)^{n_{k_1}} e^{i \phi_{k_1} \Delta}  e^{-i (\phi_{k_1} - \phi_{k_2} )S^z/2 } \nonumber \\
&+ (-1)^{n_{k_2}} e^{-i \phi_{k_1} \Delta} e^{i (\phi_{k_1} - \phi_{k_2} )S^z/2 } \nonumber \\
& + (-1)^{n_{k_1}+n_{k_2} } e^{i \phi_{k_1} \Delta} e^{-i (\phi_{k_1} + \phi_{k_2} )S^z/2 } \Big]   T_{\Delta}  | \psi_0 \rangle  .
\label{fourcircles}
\end{align}
where we introduced the shift operator
\begin{align}
T_{\Delta} =  \sum_{k=\max(0,-\Delta)}^{\min(N,N-\Delta) } |k + \Delta \rangle \langle k | .
\label{Tshiftop}
\end{align}

To verify the approximate expression (\ref{fourcircles}), we plot the Q-function of the wavefunction (\ref{step3protocol1}) which is defined as
\begin{align}
Q(\theta, \phi) = | \langle \Psi_I | \theta, \phi \rangle \rangle_3 |^2,
\label{qfuncprotocol1}
\end{align}
where the spin coherent state is on ensemble 3.  In Figure \ref{fig2} we show the Q-function for two initial states $ | \psi_0 \rangle $.  One of the initial states is chosen as a spin coherent state, and the other is a  squeezed spin coherent state, both centered at $ \theta=\pi/2, \phi = 0 $. As predicted by (\ref{fourcircles}), the effect of the measurements is to split the state into a superposition of four copies of the same state, rotated around the $ z $ axis by the angles $ \pm (\phi_{k_1} \pm' \phi_{k_2}) $. 

Taking the expectation value of the spin operators in the $ x $ and $ y $ direction with respect to the state (\ref{fourcircles}) we obtain
\begin{align}
\langle S^x \rangle & \approx \frac{1}{2} ( \cos ( \phi_{k_1} + \phi_{k_2} ) + \cos ( \phi_{k_1} - \phi_{k_2} ) ) \langle S^x \rangle_0  \\
\langle S^y \rangle & \approx \frac{1}{2} ( \cos ( \phi_{k_1} + \phi_{k_2} ) + \cos ( \phi_{k_1} - \phi_{k_2} ) )  \langle S^y \rangle_0
\label{sxsyexpression}
\end{align}
where $ \langle \vec{S} \rangle_0 $ is the expectation value with respect to the initial state $ | \psi_0 \rangle $ and we use the rotation identities
\begin{align}
    e^{-i \phi/2} S^x e^{i \phi/2} &  = \cos \phi S^x + \sin \phi S^y \nonumber \\
    e^{-i \phi/2} S^y e^{i \phi/2} & = \cos \phi S^y - \sin \phi S^y .
\end{align}
In obtaining (\ref{sxsyexpression}) we assume that the four duplicates of the state that are rotated around the $ z $ axis in (\ref{fourcircles}) are well-separated as in the case Fig. \ref{fig2}.  In this situation, the cross terms can be neglected leading to (\ref{sxsyexpression}). Substituting (\ref{phidef}) into (\ref{sxsyexpression}) and simplifying we obtain
\begin{align}
\langle S^x \rangle & \approx \frac{(2 k_1 - N) (2 k_2 - N)}{N^2}  \langle S^x \rangle_0 \label{sxsyexpression2}  \\
\langle S^y \rangle & \approx \frac{(2 k_1 - N) (2 k_2 - N) }{N^2} \langle S^y \rangle_0  .
\label{sxsyexpression3}
\end{align}

For the $ S^z $ expectation values the phase factors in (\ref{fourcircles}) do not affect the result.  The only factor that affects the result is $ T_{\Delta} $ and we obtain
\begin{align}
    \langle S^z \rangle & \approx \langle S^z\rangle_0 + 2 \Delta , 
\label{szexpect}
\end{align}
where the approximation is that the limits of the sum in (\ref{Tshiftop}) can be extended to the full range, which is valid as long as  
\begin{align}
    \langle T_\Delta^\dagger T_\Delta \rangle_0 = \sum_{k=\max(0,-\Delta)}^{\min(N,N-\Delta) } | \psi_k^{(0)}  |^2  \approx 1 . 
\end{align}
This occurs as long as the wavefunction does not have significant elements in the truncated parts of the sum.  Again, the cases we shall mainly examine are states such as spin coherent states, where the distribution tends to be peaked in $ k $, such that the approximation holds.  

We may now understand the origins for the corrections in Step 5 of the protocol in Sec. \ref{sec:protocolI}.  For the polar angle $ \theta_{\text{tel}} $, by subtracting off the constant $ 2 \Delta $ from the expectation value of $ S^z$, we obtain the original expectation value of the state $ | \psi_0 \rangle $.  For the azimuthal angle $ \phi_{\text{tel}} $, dividing (\ref{sxsyexpression3}) by (\ref{sxsyexpression2}) removes the factors associated with the measurement outcomes $ k_1, k_2 $.  To correctly recover $ \phi $ in the correct quadrant, $ \pi $ is added to the principal value of $ \tan^{-1}  $ if the sign of $ (2k_1 - N)(2k_2 - N)  $ is negative.

\subsection{Protocol II}

We now analyze Protocol II in a similar way.  The wavefunction after Step 3 of the protocol is
\begin{align}
| \Psi_{II} \rangle =   & P_{\Delta_2}^{x} P_{\Delta_1}^{z}  |\psi_0\rangle \nonumber \\
= &  \frac{1}{\sqrt{N+1}} \sum_{k=\max(0, -\Delta_1 ) }^{ \min(N, N-\Delta_1) } \sum_{k'=\max(0, -\Delta_2 ) }^{ \min(N, N-\Delta_2) }   \nonumber\\  
 & \times
\psi_{k}^{(0)}   R_{k', k} R_{k'+\Delta_2, k+\Delta_1  } |k',k'+\Delta_2\rangle^{(x)}_{12}    |k +\Delta_1  \rangle_3  .
 \label{bell_measurement_outcome}
\end{align}
The probability of obtaining the measurement outcome is 
\begin{align}
  p_{II} ( \Delta_1, \Delta_2) = & \langle \Psi_{II} | \Psi_{II} \rangle \nonumber \\
 = & \frac{1}{N+1} \sum_{k=\max(0, -\Delta_1 ) }^{ \min(N, N-\Delta_1) } \sum_{k'=\max(0, -\Delta_2 ) }^{ \min(N, N-\Delta_2) } \nonumber \\
& \times | \psi_{k}^{(0)}   R_{k', k} R_{k'+\Delta_2, k+\Delta_1  } |^2 .
\label{probprot2}
\end{align}

Using the same methods to approximate the $ R $ matrix elements as in Protocol I, we obtain the equivalent expression for the wavefunction in Protocol II
\begin{align}
& | \Psi_{II} \rangle  \approx  \frac{1}{2} \sum_{k'=\max(0, -\Delta_2 ) }^{ \min(N, N-\Delta_2) } |k', k'+\Delta_2 \rangle^{(x)}_{12} \nonumber \\
& \otimes
\Big[ e^{-i \phi_{k'} \Delta}  e^{i (\phi_{k'} + \phi_{k'+\Delta_2 } )S^z/2 } \nonumber \\
& + (-1)^{n_{k'}} e^{i \phi_{k'} \Delta_1 }  e^{-i (\phi_{k'}  - \phi_{k'+\Delta_2 } )S^z/2 } \nonumber \\
&+ (-1)^{n_{k'+\Delta_2 }} e^{-i \phi_{k'} \Delta_1 } e^{i (\phi_{k'} - \phi_{k'+\Delta_2 } )S^z/2 } \nonumber \\
& + (-1)^{n_{k'} + n_{k'+\Delta_2 } } e^{i \phi_{k'} \Delta_1 } e^{-i (\phi_{k'} + \phi_{k'+\Delta_2 }   )S^z/2 } \Big]    T_{\Delta_1}| \psi_0 \rangle_3  .
\label{fourcirclesprotocol2}
\end{align}
The main difference to the equivalent wavefunction (\ref{fourcircles}) for Protocol I is that there is entanglement between ensemble 3 and the remaining ensembles (see Fig. \ref{fig1}).  This occurs because QND measurements only partially decouple the states, when performed as a sequence in the $ z$ and $ x $ bases.  It was shown in Ref. \cite{manish2022measurement} that a long sequence of such measurements eventually results in the convergence towards an entangled state. With two measurements, there is some decoupling, but some remnant entanglement still remains.  

Evaluating the $ S^x, S^y$ expectation values with respect to the state (\ref{fourcirclesprotocol2}) we obtain
\begin{align}
    \langle S^x_3 \rangle \approx c(\Delta_2) \langle S^x \rangle_0  \label{sxprot2} \\
    \langle S^y_3 \rangle \approx c(\Delta_2) \langle S^y \rangle_0    \label{syprot2}  
\end{align}
where 
\begin{align}
c(\Delta) = \sum_{k'=\max(0, -\Delta ) }^{ \min(N, N-\Delta) } \frac{(2k'-N)(2k' +2\Delta - N)}{N^2} 
\end{align}
is an averaged version of the coefficients of the spin expectation values in (\ref{sxsyexpression2}) and (\ref{sxsyexpression3}) originating from the additional sum in (\ref{fourcirclesprotocol2}). Evaluating the sum, and factorizing we obtain
\begin{align}
c(\Delta) = \frac{2}{3} ( |\Delta | - \Delta_+ ) ( |\Delta | - \Delta_- ) ( |\Delta| - N -1 ) ,
\end{align}
where
\begin{align}
    \Delta_{\pm} = \frac{1}{2} \left( -N -1 \pm \sqrt{3N^2 + 6N + 1} \right) .
\end{align}
The function $ c(\Delta_2) $ gives negative values in the range $ \Delta_+ < | \Delta_2 | < N+1 $ and is positive otherwise.   This property of the coefficients gives the correction operation (\ref{phitel2}).  

The $ \theta_{\text{tel}} $ correction (\ref{thetatelpro2}) works in the same way as Protocol I. From the wavefunction (\ref{fourcirclesprotocol2}) we expect that
\begin{align}
\langle S^z_3 \rangle \approx \langle S^z \rangle_0 + 2 \Delta_1 ,
\end{align}
which gives the correction operation (\ref{thetatelpro2}).


\section{Performance of the teleportation protocols}
\label{sec5}

In this section, we study the performance of the protocols in Sec. \ref{sec3} by numerically evaluating the results.

\subsection{Protocol I}

\subsubsection{Single teleportation runs}
Figure \ref{fig3} shows the Bloch sphere angles of the teleported state for the various measurement outcomes using Protocol I.  The initial states $ | \psi_0 \rangle $ are chosen to be spin coherent states (Bloch sphere angles indicated by the large open circles).  For Fig. \ref{fig3}(a)(c) we omit the correction and plot the final state's Bloch sphere angles according to
\begin{align}
\theta_{\text{tel}} & = \cos^{-1} \left( \frac{\langle S^z_3 \rangle}{N} \right) \nonumber \\
\phi_{\text{tel}} & = \tan^{-1} \left( \frac{\langle S^y_3 \rangle}{\langle S^x_3 \rangle} \right) .
\label{uncorrectedspin}
\end{align}
Fig. \ref{fig3}(b)(d) includes the corrections according to (\ref{thetatelpro1}) and (\ref{phitelpro1}).  
We see that for the uncorrected case, various measurement outcomes produce teleported outcomes that have azimuthal angles $ \phi_{\text{tel}} $ that are either the same as the original value $ \phi_0 $, displaced by $ \pi $, or at $ \phi_\text{tel} = 0 $.  Without the correction operation, this produces an equal distribution in the spin variables for the $ 0 $- and $ \pi $-displaced outcomes, such that the teleportation does not transmit any information of the original state. The $ \phi_\text{tel} = 0 $ band only occurs when $ N $ is even and correspond to outcomes with either exactly $ k_1 = N/2$ or $ k_2 = N/2 $. As such they are isolated results that should become less relevant as $ N $ is increased.  For the polar angle $ \theta_{\text{tel}} $, there is a spread from the original value of $ \theta_0 $.  With the correction operations included (Fig. \ref{fig3}(b)(d)), the distribution of the states become more concentrated towards the Bloch sphere angles of the original state. Thus while the correction operation is partially effective, it also shows that the teleportation does not work perfectly deterministically.  However, we now show that the teleportation is successful in an average sense over all the outcomes.  

\subsubsection{Average error}

\begin{figure}[t]%
\includegraphics[width=\linewidth]{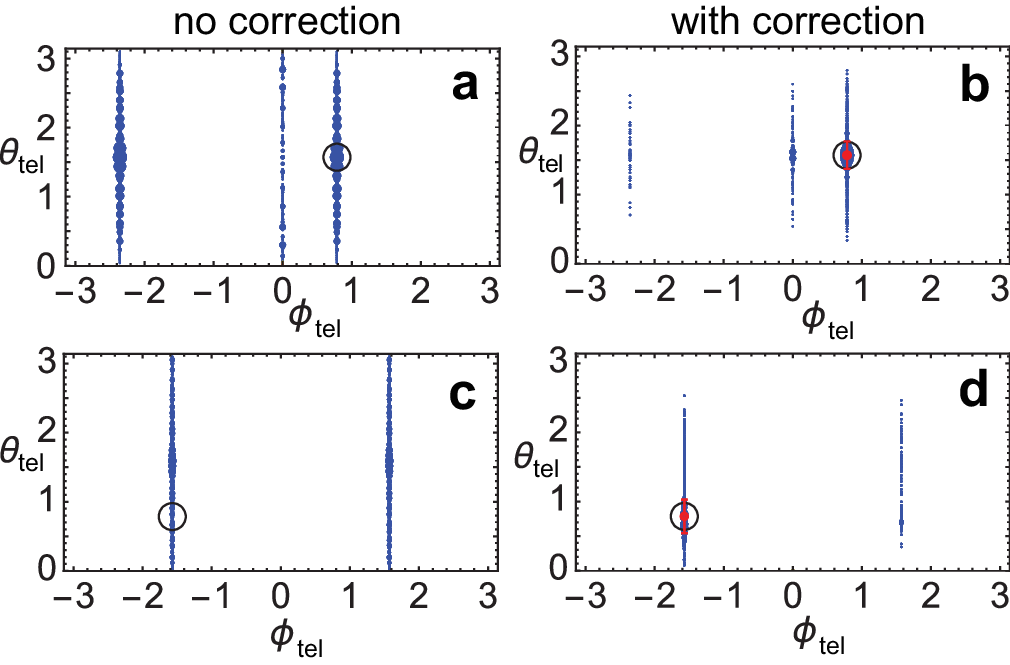}
\caption{Bloch sphere coordinates of the teleported state for Protocol I.  Each point corresponds to a measurement outcome $ (\Delta, k_1, k_2) $ and the size of the points are proportional to the probability that they occur at. For (a)(c) we omit the correction terms and plot the points according to (\ref{uncorrectedspin}).  For (b)(d) we include the correction terms as written in  (\ref{thetatelpro1}) and (\ref{phitelpro1}).  The initial state  is a spin coherent state $ | \psi_0 \rangle =| \theta_0 , \phi_0  \rangle \rangle $ with Bloch sphere angles indicated by the large open circle: (a)(b) $  \theta_0 =\pi/2, \phi_0 = \pi/4, N = 10  $, (c)(d) $ \theta_0 =\pi/4, \phi_0 = -\pi/2, N = 11 $.   The point with the error bar in (b)(d) shows the average spin value calculated with (\ref{avspin}) and (\ref{averageanglethetaphi}).  The error bars are calculated using standard deviations $ (\delta \theta_{\text{tel}}, \delta \phi_{\text{tel}}) $ following (\ref{deltatheta}). }
	\label{fig3}%
\end{figure}

We may find the average Bloch sphere coordinates according to
\begin{align}
\theta_{\text{tel}}^{\text{av}} &  = \cos^{-1} \left( \frac{S_{\text{tel}}^{z,\text{av}}}{N} \right) \nonumber \\
\phi_{\text{tel}}^{\text{av}} &  = \tan^{-1} \left( 
\frac{S_{\text{tel}}^{y,\text{av}}}{S_{\text{tel}}^{x,\text{av}}} 
\right) ,
\label{averageanglethetaphi}
\end{align}
where the averaged spin variables are 
\begin{align}
 \vec{S}_{\text{tel}}^{\text{av}} =  \sum_{\Delta, k_1, k_2 } p_I (\Delta, k_1, k_2)  \vec{S}_{\text{tel}} .
 \label{avspin}
\end{align}
Here, the teleported spin variables are given by (\ref{spintelprot1}) and the probabilities are given in (\ref{probprot1}).  In Figs. \ref{fig3}(b)(d), we see that the Bloch sphere coordinates of the average spin (point with error bar) very closely approximates the coordinates of the initial state.  We note here that the $ \phi_{\text{tel}} = 0 $ bands that are present in Figs. \ref{fig3}(a)(b) do not contribute to any error, since they correspond to $ \langle S^x_3 \rangle = \langle S^y_3 \rangle = 0 $ outcomes.  As such, they do not spoil the spin averages (\ref{avspin}).

We calculate the average error which we define as the distance between the ideal and calculated spin average value of the teleported state on the Bloch sphere. 
We define the average teleportation error as
\begin{align}
    \varepsilon_{\text{tel}} = & \frac{1}{2} \Big[
    (\sin \theta_0 \cos \phi_0 - \sin \theta_{\text{tel}}^{\text{av}} \cos \phi_{\text{tel}}^{\text{av}} )^2 \nonumber \\
    & + 
    (\sin \theta_0 \sin \phi_0 - \sin \theta_{\text{tel}}^{\text{av}} \sin \phi_{\text{tel}}^{\text{av}} )^2 \nonumber \\
   &   +  (\cos \theta_0  - \cos \theta_{\text{tel}}^{\text{av}} )^2 \Big]^{1/2}
    \label{eq:error}
\end{align}
which follows the same form as the trace distance for a qubit \cite{Nielsen:2010}. 

Evaluating the average teleportation error (\ref{eq:error}) for various $ N  $ as a function of $ \theta_0, \phi_0 $ we found that within the machine precision (i.e. $\sim 10^{-15} $) the average error is consistent with 
\begin{align}
    \varepsilon_{\text{tel}, I} = 0 .
    \label{prot1error}
\end{align}   
We do not provide a plot as it appears as numerical noise and does not give meaningful information.  The only exceptions we found are near the poles of the Bloch sphere, where the average error is at the $\sim 10^{-12} $ level, but we attribute this to the evaluation of quantities such as (\ref{uncorrectedspin}) which involve undefined quantities at the poles.

\begin{figure}[t]%
\includegraphics[width=\linewidth]{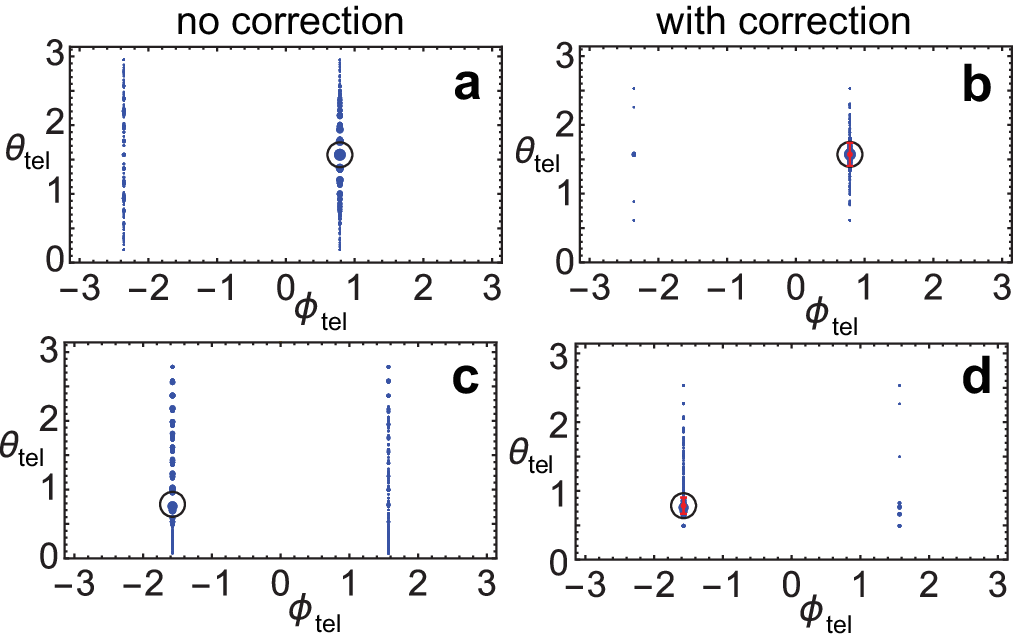}
\caption{Bloch sphere coordinates of teleported state for Protocol II.  Each point corresponds to a measurement outcome $ (\Delta_1, \Delta_2) $ and the size of the points are proportional to the probability that they occur at. For (a)(c) we omit the correction terms and plot the points according to (\ref{uncorrectedspin}).  For (b)(d) we include the correction terms as written in  (\ref{thetatelpro2}) and (\ref{phitel2}).  The initial state  is a spin coherent state $ | \psi_0 \rangle =| \theta_0 , \phi_0  \rangle \rangle $ with Bloch sphere angles indicated by the large open circle: (a)(b) $  \theta_0 =\pi/2, \phi_0 = \pi/4, N = 10  $, (c)(d) $ \theta_0 =\pi/4, \phi_0 = -\pi/2, N = 11 $.   The point with the error bar in (b)(d) shows the average spin value calculated with (\ref{avspin}) and (\ref{averageanglethetaphi}); the error bars are calculated using standard deviations $ (\delta \theta_{\text{tel}}, \delta \phi_{\text{tel}}) $ following (\ref{deltatheta}),  but using the protocol II probability (\ref{probprot2}) and spin expectations using the wavefunction (\ref{bell_measurement_outcome}). 
}	\label{fig4}%
\end{figure}

\subsubsection{Standard deviation}

While the average value closely agrees with the initial state, there is a spread of various outcomes for a particular single run of the teleportation protocol. 
To quantify the spread, we can calculate the standard deviation in the $ \theta $ and $ \phi $ directions according
\begin{align}
 \delta \theta_{\text{tel}}  = & \Big[ \sum_{\Delta, k_1, k_2 } p_I (\Delta, k_1, k_2) \left( \frac{S_{\text{tel}}^{\theta}}{N}  \right)^2 \nonumber \\
 & -  \left( \sum_{\Delta, k_1, k_2 } p_I (\Delta, k_1, k_2) \frac{S_{\text{tel}}^{\theta}}{N} \right)^2 \Big]^{1/2}
 \label{deltatheta}
\end{align}
and similarly for $ \delta \phi_{\text{tel}} $ by replacing $ {S_{\text{tel}}^{\theta}} \rightarrow {S_{\text{tel}}^{\phi}} $.  Here the probabilities are given in (\ref{probprot1}) and we have defined
\begin{align}
S_{\text{tel}}^{\theta} & = \cos \theta_0 \cos \phi_0 S^x_{\text{tel}} + \cos \theta_0 \sin \phi_0 S^y_{\text{tel}}
- \sin \theta_0 S^z_{\text{tel}} \nonumber \\
S_{\text{tel}}^{\phi} & = - \sin \phi_0 S^x_{\text{tel}}  + \cos \phi_0  S^y_{\text{tel}}
\label{thetaphispins}
\end{align}
according to the formula for unit vectors in spherical coordinates. From the error bars in Figs. \ref{fig3}(b)(d) we see that there is negligible error in the $ \phi $ direction, which is as expected since both the $0$- and $ \pi$- shifted distributions are very narrow.   The $ \pi$-shifted distribution does not contribute to the $ \phi$ spread since it is exactly on the opposite side of the Bloch sphere. We find that for all parameters our results are consistent with
\begin{align}
\delta \phi_{\text{tel}, I} = 0
\end{align}
to within numerical precision.  The dominant spread is thus in the $ \theta $-direction.

\begin{figure}[t]%
\includegraphics[width=\linewidth]{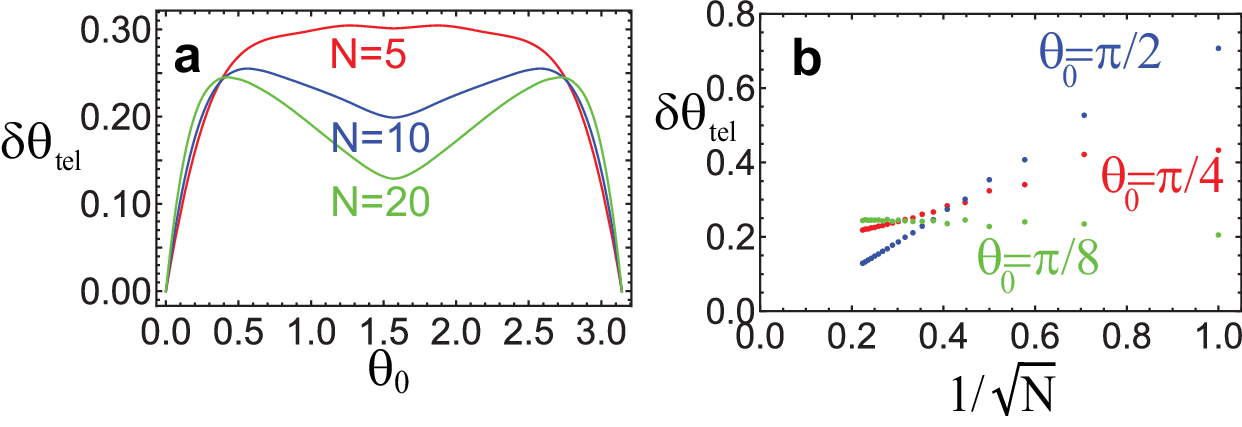}
\caption{The standard deviation (\ref{deltatheta}) of the teleported spin variable $ S^{\theta}_{\text{tel}} $ for Protocol I calculated with (\ref{thetaphispins}),  (\ref{spintelprot1}), and (\ref{probprot1}).  
 For all plots we use $ \phi_0 = \pi/4 $ to calculate the points, but the results are independent of $ \phi_0 $.  }
	\label{fig5}%
\end{figure}

In Fig. \ref{fig5}(a), the standard deviation of the teleported states in the $ \theta $ direction is shown as a function of $ \theta_0 $.  The curves are independent of $ \phi_0 $, hence we consider the $ \phi_0 =\pi/4 $ as a representative case.  We see that near the poles of the Bloch sphere, there is negligible spread and the dominant spread is for the intermediate angles.  The scaling with $ N $ is shown in Fig. \ref{fig5}(b).  We see that for the $ \theta_0 = \pi/2 $ the standard deviation approaches zero, achieving ideal teleportation in the large $ N $ limit.  For $ \theta_0 $ not at the poles nor the equator the scaling suggests that there is some remaining variance in the large $ N $ limit, at the level of $ \delta \theta \sim 0.1 $.

In summary,  Protocol I  does not involve any kind of systematic deviation when looking at averaged quantities, but on a single run of the teleportation protocol there can be deviations.  These deviations are only in the $ \theta $ direction and tend to be fairly modest, at the $ \delta \theta \sim 0.1 $ value for large $ N $.

\subsection{Protocol II}

\subsubsection{Single teleportation runs}

Fig. \ref{fig4} shows the Bloch sphere angles of the teleported state for the various measurement outcomes.  We use a spin coherent state as the initial state and plot the angles according to (\ref{uncorrectedspin}), using the wavefunction (\ref{bell_measurement_outcome}).  Similarly to Fig. \ref{fig3} for Protocol I,  Fig. \ref{fig4}(a)(c) omits the correction terms and Fig. \ref{fig4}(b)(d) shows the results including the correction terms (\ref{spintelprot2}).  We see largely the same behavior as Protocol I (Fig. \ref{fig3}).  Without the correction there is an equal distribution around $ \phi_0 $ and $ \phi_0 + \pi $.  Including the correction largely eliminates the outcomes with $ \phi_0 + \pi $.  This appears to be slightly better than the Protocol I correction scheme which still has a noticeable band at $ \phi_0 + \pi $.  One major difference to Protocol I is that for even $ N $ there is no band at $ \phi_{\text{tel}} = 0 $.  This is due to the different nature of the correction scheme, where it is less likely that $ \Delta_2 = \Delta_+ $ exactly, due to the integer values that $ \Delta_2 $ takes, while  $  \Delta_+ $ may not be integer-valued.

\subsubsection{Average error}

We evaluate the average teleportation error (\ref{eq:error}) for various $N $ as a function of $ \theta_0, \phi_0 $.  The average spin is evaluated with (\ref{avspin}) but using the Protocol II probability (\ref{probprot2}) and the teleported spin variables (\ref{spintelprot2}).  we find that our results are consistent with 
\begin{align}
    \varepsilon_{\text{tel}, II} = 0 
       \label{prot2error}
\end{align}   
to within machine precision (i.e. $ \sim 10^{-15} $), except near the poles of the Bloch sphere as was the case with Protocol I. We do not provide a plot as it appears as numerical noise and does not give meaningful information. 

\begin{figure}[t]%
\includegraphics[width=\linewidth]{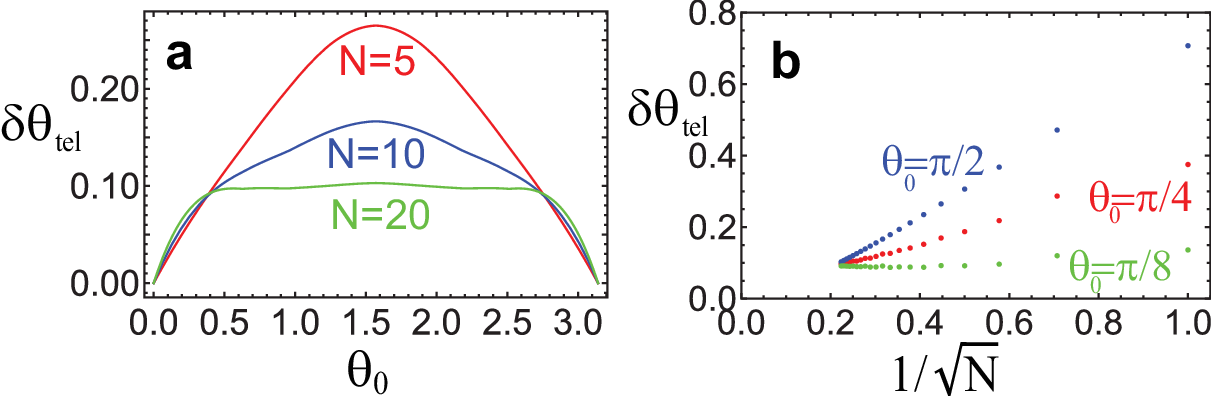}
\caption{The standard deviation (\ref{deltatheta}) of the teleported spin variable $ S^{\theta}_{\text{tel}} $ for Protocol II calculated (\ref{thetaphispins}), (\ref{probprot2}), and (\ref{spintelprot2}).  For all plots we use $ \phi_0 = \pi/4 $ to calculate the points, but the results are independent of $ \phi_0 $.  }
	\label{fig6}%
\end{figure}

\subsubsection{Standard deviation}

We evaluate the standard deviation of the various measurement outcomes.  
This is evaluated in a similar way as (\ref{deltatheta}) but with the Protocol II probabilities (\ref{probprot2}) and using the transformed spin variables (\ref{thetaphispins}) with (\ref{spintelprot2}). In the $\phi$ direction, we again find that to numerical precision
\begin{align}
\delta \phi_{\text{tel}, II} = 0 .
\end{align}
Figure \ref{fig6} shows the standard deviation of the spin variable in the $ \theta $ direction. As with Protocol I, there is no dependence with $ \phi_0 $ hence we consider the dependence with $ \theta_0 $. The standard deviation tends to give slightly smaller values in comparison to Protocol I, with a dependence that tends to be more peaked at the equator of the Bloch sphere (Fig. \ref{fig6}(a)).  Examination of the scaling with $ N $ (Fig. \ref{fig6}(b)) shows that again for large $ N $ the equatorial states extrapolate to small spread, whereas the states between the poles and equator appear to extrapolate to a non-zero value, at the level $ \delta \theta \sim 0.1 $.

In summary, we find Protocol II does not have any systematic deviation of averaged quantities.  For single runs of the protocol, there are deviations at the $ \delta \theta \sim 0.1 $ level for large $ N $ but none in the $ \phi $ direction.

\section{Comparison to classical bounds and decoherence}
\label{sec6}

Our teleportation protocols only achieve sending the total spin expectation values of the state $ | \psi_0 \rangle $ and not the state itself.  The reason for this is visible in Fig. \ref{fig2} --- the state that is transmitted involves a four-fold duplicate of the original state.  This raises the question of exactly how quantum the protocol is.  Is there a classical strategy that could match or outperform the scheme?  These questions are important also from the perspective of proving that a non-trivial quantum operation has been performed.  Typically, experimental demonstrations of quantum teleportation are accompanied by classical bounds in the fidelity to show that a non-classical operation has been performed. 

A purely classical strategy involves Alice measuring the spin expectation value $ \langle \vec{S} \rangle_0 $ using only one copy of the state $ | \psi_0 \rangle$.  She then classically transmits her best estimate of the spin expectation value to Bob.  For spin coherent states, we may use the quantum state estimation bounds derived by Massar and Popescu, where the optimal fidelity of estimating $ N $ copies of a qubit is $ F = (N+1)/(N+2) $.  Converting this to a qubit trace distance \cite{fuchs1999cryptographic} defined as in (\ref{eq:error}), we have \cite{pyrkov2014full}
\begin{align}
\varepsilon_{\text{QSE}} \ge \sqrt{1 - F} = \frac{1}{\sqrt{N+2}} ,
\end{align}
which represents the lowest error that can be attained using the classical strategy.  Comparing this to the average error of Protocols I and II given as (\ref{prot1error}) and (\ref{prot2error}), we find that in the ideal case
\begin{align} 
    \varepsilon_{\text{tel}} <  \varepsilon_{\text{QSE}} 
\end{align}
for both Protocols I and II, showing that the teleportation exceeds the performance of a purely classical strategy.  As shown in the next section, decoherence will increase the average error, which may lift the teleportation error above the classical bound.

\begin{figure}[t]%
\includegraphics[width=\linewidth]{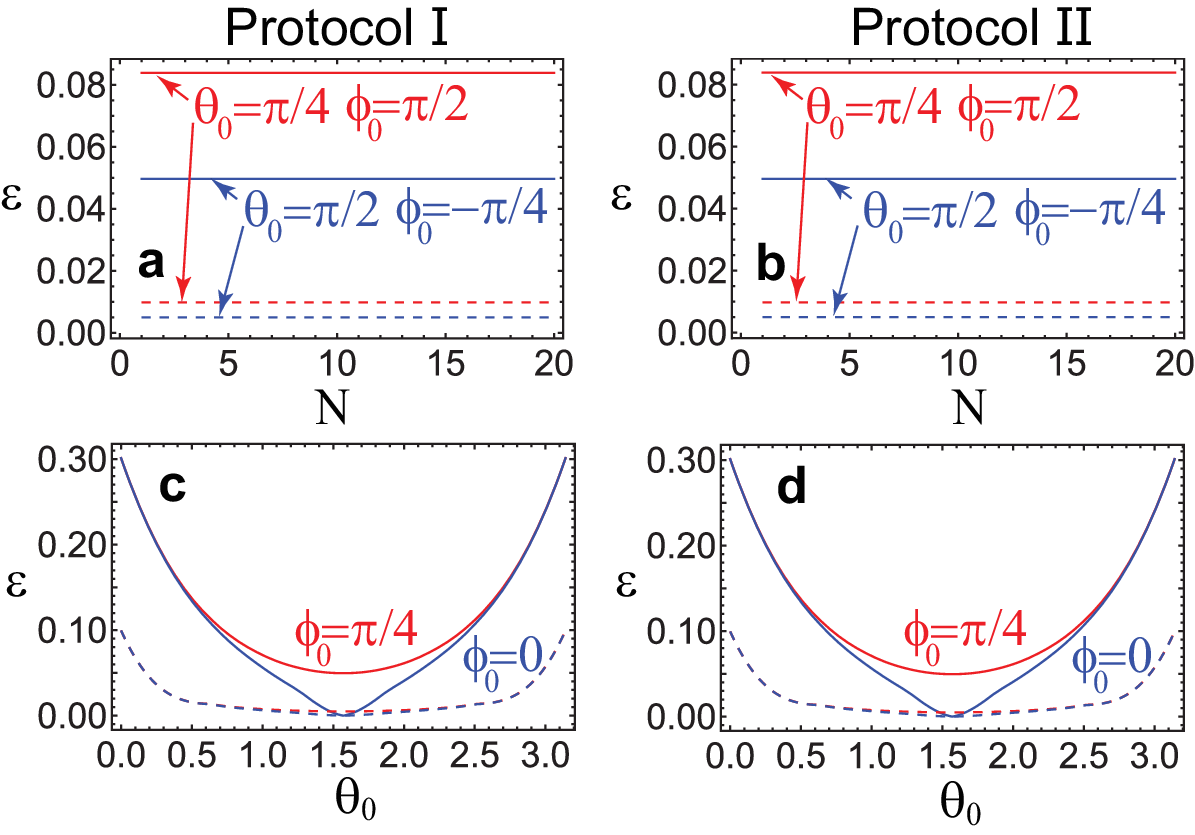}
\caption{Average error in the presence of decoherence for (a)(c) Protocol I and (b)(d) Protocol II.  The dependence with $ N $ for the initial state with the angles as marked are shown for (a)(b).  Subplots (c)(d) show the angular dependence for $ N = 10$. Dashed lines show the results for $ \gamma t = 0.01$ while solid lines show the results for $ \gamma t = 0.1 $ for all plots. }
	\label{fig7}%
\end{figure}

\section{Decoherence}
\label{sec7}
An important consideration in the context of macroscopic teleportation is decoherence.  Typically macroscopic quantum states (e.g. Schrodinger cat states) decohere extremely quickly, rendering them impossible to realize them experimentally \cite{zurek2003decoherence}.  In the case of atomic ensembles, the primary forms of decoherence that take place are in the form of atomic loss and dephasing \cite{byrnes2015macroscopic}. 
The most sensitive part of the protocol to decoherence is the preparation of the maximally entangled state (\ref{mmesstate}) and also the final states (\ref{step3protocol1}) and (\ref{bell_measurement_outcome}) which involve Schrodinger cat-like states.  Such states involve macroscopic superpositions and are potentially susceptible to decoherence that scale unfavorably with $ N $.  Therefore, on first glance it may appear that our protocol cannot work effectively when decoherence is included.  In this section, we show that in fact our protocol is rather robust under decoherence. 

To understand the effect of decoherence, we apply dephasing according to 
\begin{align}
\frac{d \rho}{dt} = -\frac{\gamma}{2} \sum_{l=1}^3 \Big[ (S^x_l)^2 \rho - S^x_l \rho S^x_l + \rho (S^x_l)^2 \Big] ,
\label{lindblad}
\end{align}
which is applied to the initial state (\ref{initialstate}) and also after the measurements are made (after Steps 1 and 3 in both Protocols I and II).  We then evaluate the expectation values and evaluate the average error according to (\ref{eq:error}). We apply dephasing in the $ x $ direction because in Step 2 of the protocol we apply a QND projection in the $ z $ direction.  Dephasing in the $ z $ direction commutes with dephasing and gives a more trivial dependence.  The details of the calculation are deferred to Appendix \ref{app:deco}.  

Figure \ref{fig7}(a)(b) shows the dependence of the average error (\ref{eq:error}) with $ N $ for various initial states and decoherence rates.  We see that in all cases the average error has no dependence on $ N $.  There is very little difference between the Protocol I and II cases.  The reason for the independence with $ N $ may be understood as follows.  
After the measurement steps in the teleportation protocols, states such as shown in Fig. \ref{fig1} are generated, which are Schrodinger cat-like states.  These are typically highly susceptible to decoherence in terms of the states themselves, but spin expectations are generally unaffected.  For example, consider the Schrodinger cat state
\begin{align}
| \xi \rangle \approx \frac{1}{\sqrt{2}} ( | \theta, \phi \rangle \rangle + | \theta', \phi' \rangle \rangle )
\end{align}
where we took the coordinates $ ( \theta, \phi) $ and $ ( \theta', \phi') $ sufficiently distant such that $ \langle \langle \theta', \phi'  | \theta, \phi \rangle \rangle \approx 0 $.  Under decoherence such as state very quickly collapses to a incoherent mixture
\begin{align}
\rho \approx \frac{1}{2} ( | \theta, \phi \rangle \rangle \langle \langle \theta, \phi | + | \theta' , \phi' \rangle \rangle \langle \langle \theta', \phi' | ) . 
\end{align}
Evaluating spin expectation values, the only difference between the cat state and the incoherent mixture is the presence of cross terms
\begin{align}
\langle \xi | \vec{S} | \xi \rangle - \text{Tr} ( \rho \vec{S} ) = 
\text{Re} ( \langle \langle \theta, \phi | \vec{S} | \theta' , \phi' \rangle \rangle ) .
\label{differencecatmixture}
\end{align}
Since the spin operators only cause single number state transitions $ | k \rangle \rightarrow | k \pm 1 \rangle $, for well separated cat states, the difference (\ref{differencecatmixture}) is exponentially suppressed \cite{timquantumoptics2020}.  In deriving the expectation values in Sec. \ref{sec4} we have approximated that such cross-terms are negligible.  In the presence of decoherence, neglecting such terms become exact.  For these reasons, the spin expectation values of the teleportation should be unaffected by the decoherence of the cat states generated after the measurement steps of the teleportation protocol.  

In our calculations we also apply decoherence to the initial state (\ref{initialstate}).  From the results in Fig. \ref{fig7}(a)(b), we conclude that there is no sensitivity with $ N $, and only has the overall effect of dampening the spin expectation values of the teleported state in the $ y $ and $ z $ directions.  The stability of the maximally entangled state under dephasing was examined in works such as Ref. \cite{byrnes2013fractality,Gao_2022,shuai2023,PhysRevA.106.L051701} 
and was found to be more robust than states such as Schrodinger cat states, since they do not directly involve macroscopically different superpositions.  Since the technique to prepare it (Ref. \cite{manish2023maximallyentangled}) is a measurement based method, we expect it to be more robust as the projections act to project the state into the desired spin sector. We therefore expect that as long as the teleportation protocols are executed within a suitable decoherence window, these can be performed with the high fidelity value. 

In Fig. \ref{fig7}(c)(d) we also show the $ \theta_0 $ dependence of the average error for various $ \phi_0 $ values at a fixed $ N $.  We see that the largest effect of the decoherence occurs for the states at the poles of the Bloch sphere.  This is not surprising as we apply dephasing in the $ x $ direction, and spin coherent states polarized in the $ z $ direction are most susceptible.  As the state approaches the $ x $ direction, the state becomes less affected, since the dephasing only creates a global phase.  There is again very little difference between the effect on Protocol I and II, since both effectively teleport the initial state.   

In summary, we expect our teleportation protocols are highly robust in the presence of decoherence.  The primary reason for this is that the teleported variables are spin expectation values, rather than the full quantum states themselves. We expect similar results for other types of decoherence.  For example, for atomic loss, the total spin operator  merely results in an amplitude damping of the total spin.

\section{Experimental implementation}
\label{sec8}

We now briefly describe the experimental requirements necessary for realizing our teleportation protocols.  Our protocols have been designed with experimental methods that have been demonstrated in the past, hence we believe that our approach should be experimentally accessible. 

The qubit ensembles can be realized with neutral gas ensembles, trapped in glass cells, or with multiple optical/magnetic traps such as with an atom chip.  Each atom has two populated internal states.  A common choice for the internal states are hyperfine ground states, such as the $ F= 1, m_F = - 1 $ and $ F = 2, m_F = 1 $ states in the case of $^{87}$Rb \cite{pezze2018quantum}.  Either hot or cold atomic ensembles or BECs may be used, due to the equivalence of the total spin operators in the symmetric subspace \cite{timquantumoptics2020}. Three ensembles are required to perform the quantum teleportation, involving either multiple atomic cells or with an atom chip with multiple traps.  For the preparation of the initial state, coherent microwave frequency radiation can prepare spin coherent states, as is routinely performed with atomic ensembles \cite{pezze2018quantum}.  

Both Protocol I and II rely on using the QND measurements as described in Appendix \ref{app:qnd}. QND measurements have already been proven to be an effective method for generating entanglement between atomic ensembles \cite{kuzmich2000generation, julsgaard2001experimental,hammerer2010quantum}. The Mach-Zehnder configuration as shown in Fig. \ref{Figapp0} is an equivalent way of performing the same task, which measures the relative $ S^z_1 - S^z_2 $ eigenvalue \cite{aristizabal2021quantum}.  The details of such measurements in a POVM framework is given in Refs.  \cite{ilo2023measurement,manish2022measurement}.  Numerous works have shown that the type of entanglement that is generated is robust against various decoherence sources such as photon loss and spontaneous emission \cite{ilo2024quantum,Gao_2022,shuai2023}. The main challenges will be to precisely realize the projection operator (\ref{projzbasis_singlepeak}), which ideally gives a readout at the single atom level.  The outcomes $ \Delta $ corresponds to the eigenvalues of $ (S^z_1 - S^z_2)/2 $ which is deduced from the photon number outcomes.  A sufficiently large photon amplitude, combined with an accurate estimation of the photon intensity should give a good estimate on $ \Delta $, but to our knowledge this has not been experimentally realized.   With an accurate QND measurement, the maximally entangled state can be deterministically prepared as discussed in Sec. \ref{sec:mmes}.  

Finally, in order to read out the state of the spins on ensemble 3, and also to perform the projective measurements in Protocol I, the single ensemble projective measurement (\ref{projz}) is required.  This is routinely performed in atomic ensemble experiments, for example using absorption imaging \cite{bohi2009coherent,riedel2010atom}.  A precise readout is required for both a precise estimation of the spin $ \langle \vec{S} \rangle $ and to perform the correction operations in Protocol I. Readout at near-single atom resolution has been demonstrated in Refs. \cite{bucker2009single,PhysRevLett.111.253001,PhysRevA.97.063613,alberti2016super}.

While we have emphasized the readout at the single atom level above, we note that our protocols are {\it not} in fact extremely sensitive to readout error, both in the final spin readout nor the correction operations.  For example, the spin readout $ \langle S^z \rangle $ is evaluated by counting the difference between the $ a $ and $ b $ qubit states, and errors in counting these only affect the Bloch sphere angles $ \theta_{\text{tel}}, \phi_{\text{tel}} $ unless they are of magnitude $ \sim N $.  Similarly, the correction operations such as $ H(k_{1,2} - N/2) $ and $ H(|\Delta_2| - \Delta_+ ) $ are only significantly affected when the errors in the readouts $ k_{1,2} $ and $ \Delta_2 $ are at the $ \sim N $ level.  This is rather different, for example, to if the correction operations involved the parity of $ k_{1,2} $ and $ \Delta_2 $, which are very sensitive to the readout error.

\section{Summary and Conclusions}
\label{sec9}

We have introduced two teleportation protocols to transmit an unknown spin expectation values using entanglement and classical communication based on qubit ensembles.  One of the main features of the protocol is that it does not use any operations requiring microscopic control of the qubits within each ensemble.  The only operations that are required consist of collective spin rotations and measurements of each ensemble, and entangling QND measurements between two ensembles.  An arbitrary spin pointing in any direction of the Bloch sphere can be teleported, hence the protocol goes beyond the conventional Holstein-Primakoff approximation for performing continuous variable teleportation.  We have introduced two versions of the teleportation scheme (Protocols I and II) since they have a similar performance albeit with small differences in the general behavior.  Which protocol is chosen depends entirely on which is more convenient for the implementation. 

In terms of the performance, we point out that the protocols are not perfectly deterministic in the sense that every measurement outcome yields a spin expectation that is precisely in the same spin direction as the initial state.  However, when averaged over measurements (including the classically communicated correction), the averaged spin remarkably gives {\it perfect} transmission, for the case of a spin coherent state.  While we are able to perform analytical calculations of the resulting state, this has involved approximations, notably of the matrix elements $ \langle k | k' \rangle^{(x)} $.  In general such factors deviate the state from a simple state such as a spin coherent state (see for e.g. Eq. (\ref{step3protocol1})) hence it is not obvious how the spin expectations are so perfectly preserved.  This may point to an alternative way of analyzing the resulting wavefunctions without resorting to the approximations which we have implemented.  We note that we mainly considered initial states that are spin coherent states, and the protocol could be used for other types of states.  For other states we expect that the perfect transmission may be adversely affected since the way they are formulated in Sec. \ref{sec3}, only angular variables are transmitted, and some states may not be on the surface of the Bloch sphere as is implicitly assumed here.  

While our protocol is not perfectly deterministic, they are quasi-deterministic in the sense that the standard deviation of the states distributed around the Bloch sphere is rather tightly distributed.  The distribution only occurs in the $ \theta $ direction, and in the $\phi $ direction there is no variance since points with errors occur with a phase difference of $ \pi $.  To derive the correction steps in Protocol I and II, we have used the analytical approximations, and this is successful in picking up a majority of the states. But as can be seen from Figs. \ref{fig3} and \ref{fig4} some outcomes are not corrected perfectly.  Hence there is still scope for improvement to use the correction protocol in order to better approximate a deterministic scheme. 

We have tested the robustness of our protocol under decoherence and found that it is rather robust.  For the dephasing model that we considered, there is no dependence on the ensemble size $N $.  This is important since in teleporting macroscopic states, one potential point of failure is that coherence rapidly degrades for large systems, such as in Schrodinger cat states.  Our protocol has been designed with this in mind, where more robust states such as spin coherent states and the maximally entangled state $ | \Phi \rangle \rangle $ are used.  We have analyzed the effect of decoherence in previous works \cite{byrnes2013fractality,Gao_2022,shuai2023,manish2021remote,manish2023maximallyentangled} and know that such states are generally rather robust.  Teleporting the total spin variables (rather than the full state) is also a key factor that allows for the protocol to be successful in the presence of decoherence, since spin operators tend to only decohere at the same rate as that for the corresponding qubit variables.  Our protocol is however not extremely sensitive to readout errors since only spin averages and correction operations in two broad regions of the measurement outcome space are required.  We therefore expect that our protocols should be experimentally accessible.


\begin{acknowledgments}
This work is supported by the SMEC Scientific Research Innovation Project (2023ZKZD55); the Science and Technology Commission of Shanghai Municipality (22ZR1444600); the NYU Shanghai Boost Fund; the China Foreign Experts Program (G2021013002L); the NYU-ECNU Institute of Physics at NYU Shanghai; the NYU Shanghai Major-Grants Seed Fund; and Tamkeen under the NYU Abu Dhabi Research Institute grant CG008. M.C. acknowledges the funding support from the Department of Science and Technology, Government of India, through the I-HUB Quantum Technology Foundation, IISER Pune, India and the FWO and the F.R.S.-FNRS as part of the Excellence of Science program (EOS project 40007526) at University of Liège, Belgium.
\end{acknowledgments}

\appendix

\section{Projection operators for QND measurements}
\label{app:qnd}

Here we summarize the theory of QND measurements in the positive operator valued measure (POVM) formalism \cite{aristizabal2021quantum,manish2022measurement,ilo2023measurement}. 

\begin{figure}[t]%
	\includegraphics[width=\linewidth]{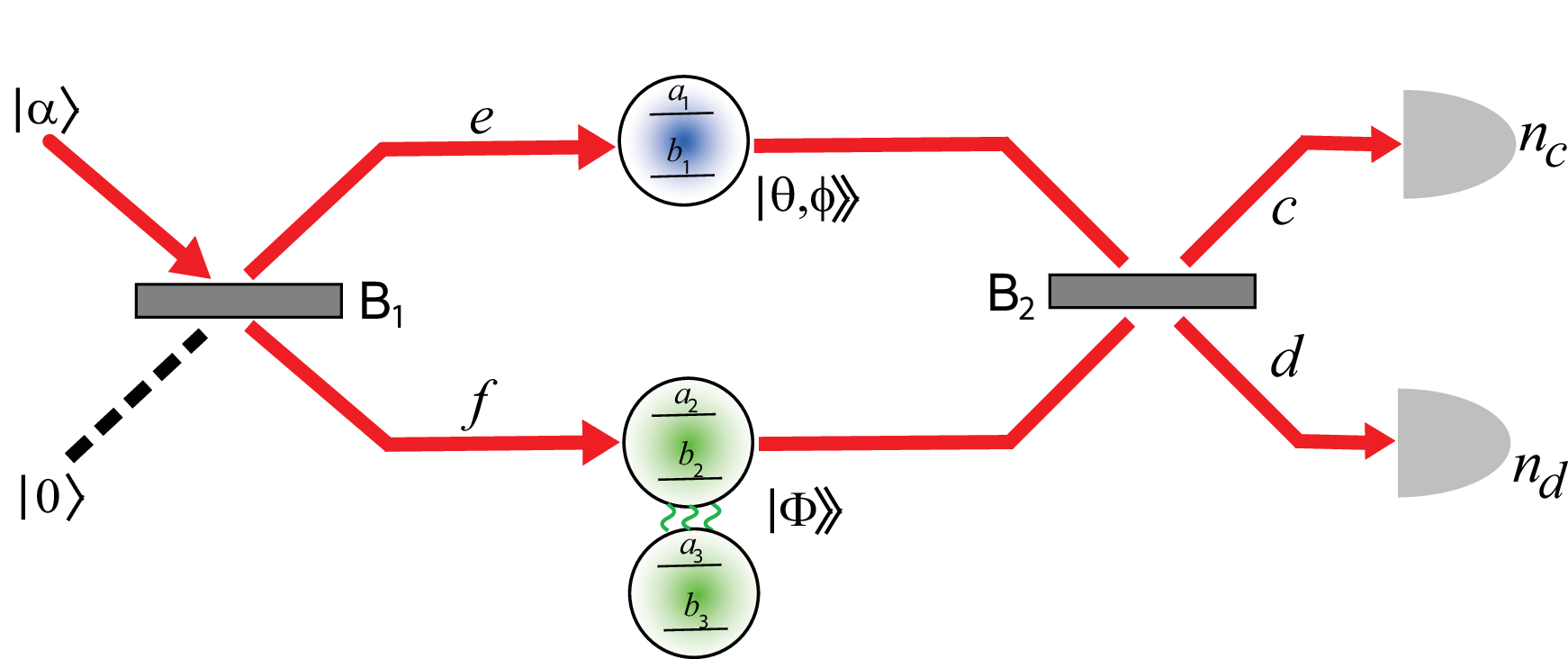}
	\caption{Entanglement generating QND measurements between qubit ensembles.  
 Coherent light $|\alpha\rangle$ interacts with ensembles via the QND Hamiltonian interaction (\ref{interhamiltonian}) arranged in a Mach-Zehnder configuration. After the second beam splitter $B_2$, the photon modes are detected yielding counts $n_c,n_d$.  This entangles ensemble 1 with ensemble 2.  }
	\label{Figapp0}%
\end{figure}

The basic setup for a QND measurement is shown in Fig. \ref{Figapp0}.  Coherent light is used to perform an indirect measurement of the atomic ensembles arranged in a Mach-Zehnder configuration. The interaction between photons and atoms is governed by the Hamiltonian \cite{kuzmich2000generation}, 
\begin{align}
	H = \kappa(S^z_{1}-S^z_{2}+ 2N )J^z ,
	\label{interhamiltonian}
\end{align}
where $\kappa$ is the coupling constant and $J^z = e^\dagger e - f^\dagger f $ is the Stokes operator for the two optical modes $e,f$ that enter into each arm of the interferometer.  After interacting with the atoms, the photonic modes are interfered with a beam splitter, giving rise to new modes $c,d$ and the photons are detected by the detectors with counts $n_c,n_d$ respectively.
%
%
%

The QND measurement between two ensembles can be described in terms of a Positive Operator Valued Measure (POVM) as \cite{manish2022measurement}
\begin{align}
M_{n_c n_d}(\tau) =\sum_{k_1,k_2 = 0}	C_{n_c,n_d}[(k_1-k_2+N)\tau]|k_1,  k_2\rangle \langle k_1 , k_2| ,
\label{povmdef}
\end{align}
where $ |k_1, k_2 \rangle = |k_1 \rangle \otimes |k_2 \rangle $ and 
the modulating function is defined as
\begin{align}
	C_{n_c,n_d}(\chi) = \frac{\alpha^{n_c+n_d}e^{-|\alpha|^2/2}}{\sqrt{n_c!n_d!}}\cos^{n_c}(\chi)\sin^{n_d}(\chi),
 \label{cfunc}
\end{align}
and $\tau = \kappa t$ is the interaction time. The resulting state after the measurement is
\begin{align}
 M_{n_c n_d}(\tau) | \psi \rangle   = \sum_{k_1, k_2}  \psi_{k_1 k_2}  C_{n_c, n_d} [ (k_1 - k_2+N) \tau] | k_1, k_2 \rangle .
\label{psichange}
\end{align}
where $ \psi_{k_1, k_2} = \langle k_1, k_2 | \psi\rangle $.
Here we see that the initial wave function is modulated by an extra factor of $C_{n_c, n_d} [ (k_1 - k_2+N) \tau] $.

For large photon number $ |\alpha|^2 \gg 1 $ and short interaction times $ \tau \sim 1/N $, the modulating function $C_{n_c, n_d} $ takes a Gaussian form \cite{aristizabal2021quantum} and is sharply peaked at
%
\begin{align}
\sin^2[(k_1-k_2+N)\tau] = \frac{n_d}{n_c+n_d}.
\label{deltancnd}
\end{align}
This is the mechanism that results in the measurement-induced generation of entanglement since it enforces the argument of the modulating function to be a constant value, i.e. $ (k_1 - k_2) \tau = \text{const} $.  This is equivalent to a measurement of the operator $ S^z_1 -S^z_2 $.  For smaller photon counts the modulating function becomes broader and corresponds to a weak measurement.


Taking the interaction time $ \tau  = \pi/4N $ and assuming $| \alpha \tau |^2 \gg 1 $,  as defined in Ref. \cite{manish2022measurement}, we may then approximate the POVM (\ref{povmdef}) as a projection operator (\ref{projzbasis_singlepeak}) according to
\begin{align}
M_{n_c n_d} ( \tau= \frac{\pi}{4N}  ) \approx  
P_\Delta^{z} ,
\end{align}
where the projections $ \Delta = k_1-k_2 $ and photonic measurements $ n_c, n_d $ are related according to (\ref{deltancnd}).

\section{Maximally entangled state preparation}
\label{app:maxent}

The protocol involves repeated measurements of QND projection operator (\ref{projzbasis_singlepeak}) and (\ref{projzbasis_singlepeak_x}) on two ensembles.  The method takes advantage of the fact that the state (\ref{mmesstate}) is a simultaneous eigenstate of the projection operator with probability 1 
\begin{align}
  P_0^z |\Phi \rangle\rangle   = P_0^x |\Phi \rangle\rangle  = |\Phi \rangle\rangle .
\end{align}
This means that once the state (\ref{mmesstate}) is obtained, the QND measurements no longer affect the state, and results in a fixed point of the repeated measurements.  This is however only true for the measurement outcome $ \Delta = 0 $ in both the $ x $ and $ z $ basis.  Since quantum measurements are intrinsically random, we must account for the fact that $ \Delta \ne 0 $ may be obtained.  To overcome this an adaptive measurement sequence is performed:
\begin{align}
    T_{\vec{\Delta}}^z = \prod_{j=1}^L U_{\Delta_j} P_{\Delta_j}^{z}
\end{align}
where $ \vec{\Delta} = ( \Delta_1,\Delta_2, \dots, \Delta_L) $ is the measurement sequence where the last measurement value is $ \Delta_L = 0 $.  A suitable adaptive unitary is 
\begin{align}
U_{\Delta} = e^{i S^y \pi \Delta / 2N } \otimes I .
\end{align}
Then by applying such a sequence in both the $ z $ and $ x $ basis, an arbitrary state is driven towards the maximally entangled state
\begin{align}
\prod_{r=1}^M (T_{\vec{\Delta}^x_r }^x T_{\vec{\Delta}^z_r}^z ) | \psi_0 \rangle \rightarrow |\Phi \rangle\rangle  ,
\end{align}
where $ T_{\vec{\Delta}}^x =  e^{- i (S^x_1+ S^x_2) \pi/4 } T_{\vec{\Delta}}^z  e^{i (S^x_1+ S^x_2) \pi/4 }$. 

In Ref. \cite{manish2023maximallyentangled} a slightly different QND measurement was used but the above sequence (\ref{projzbasis_singlepeak}) and (\ref{projzbasis_singlepeak_x}) is equally effective at producing the maximally entangled state.  

In this paper, we primarily consider the case where Alice and Bob are not separated by large distances. In principle, generating entanglement separated by large distances is not a fundamental obstacle to the above scheme as it relies only on QND measurements and local spin rotations.  QND measurements are performed using bright coherent light \cite{aristizabal2021quantum}, which is not as sensitive as alternative entanglement generation approaches relying on single photon transmission, for example.  Using a polarization encoding of the modes, long-distance interferometry should be possible \cite{ilo2023measurement}, realizing long-distance entanglement.  For other types of entanglement, other approaches have also been developed using more conventional quantum repeater methods \cite{pyrkov2024quantum}.

\section{Matrix elements of the spin rotation operator}
\label{app:prob}

The number states $|k\rangle $ are eigenstates of the $ S^z $ spin operator.  
Number states that are eigenstates of spins in other directions can be obtained by applying spin rotations (\ref{oneensrot}).  The matrix elements of the $S^y$ rotation are given by \cite{timquantumoptics2020}
\begin{align}
   &\langle k'|e^{-iS^y \theta/2}|k\rangle =  \sqrt{k!(N-k)!k'!(N-k')!}  \nonumber \\ &  \times    \sum_{n=\text{max}(k'-k,0)}^{\text{min}(k',N-k)}\frac{(-1)^n\cos^{k'-k+N-2n}(\theta/2)\sin^{2n+k-k'}(\theta/2)}{(k'-n)!(N-k-n)!n!(k-k'+n)!} .
      \label{matrixelem}
\end{align}
Using the definition (\ref{kxbasisdef}) and the above formula for $ \theta = \pi/2 $ we have
\begin{align}
&  R_{k',k}    = \langle k'|^{(x)} | k \rangle =  \langle k| k' \rangle^{(x)} \nonumber = \langle k|e^{-iS^y \pi/4}|k'\rangle \nonumber \\
 & =  \sqrt{\frac{k!(N-k)!k'!(N-k')!}{2^N}}  \nonumber \\ 
 &  \times    \sum_{n=\text{max}(k-k',0)}^{\text{min}(k,N-k')}\frac{(-1)^{n}}{(k-n)!(N-k'-n)!n!(k'-k+n)!} .
 \label{rmatrix}
\end{align}

\section{Harmonic oscillator eigenstate approximation of the inner product of number states in the $ x $ and $ z $ basis}
\label{app:kxkzapprox}

\begin{figure}[t]%
\includegraphics[width=\linewidth]{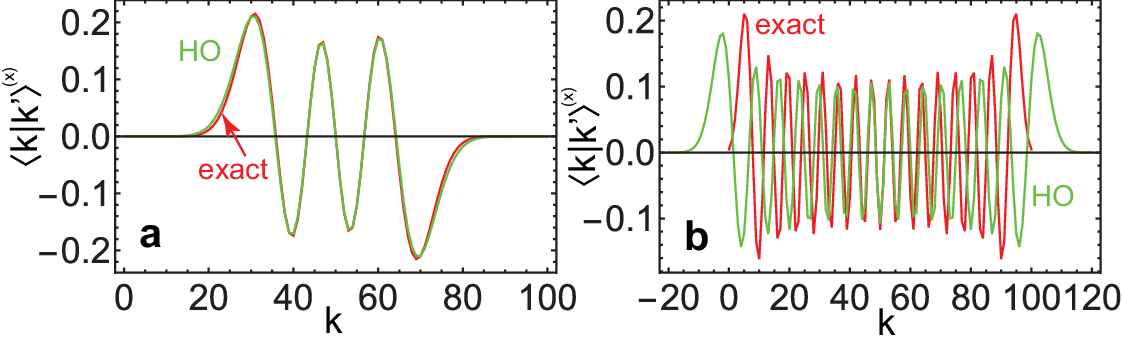}
	\caption{Approximating the inner product $ \langle k | k' \rangle^{(x)} $ with the harmonic oscillator wavefunction (labeled as HO).  The inner product is calculated using (\ref{rmatrix}) and the harmonic oscillator wavefunction (\ref{finalapprox}) is plotted.   Parameters chosen are (a) $ k' = 95 $; (b) $ k' =70 $.  We use $ N  = 100 $ for all plots.  }
	\label{fig:app1}%
\end{figure}

In this section we approximate the matrix elements (\ref{rmatrix}).  Our basic approach is to use the Holstein-Primakoff approximation to relate the spin operators to position and momentum variables.  

Consider the state $ |k' \rangle^{(x)} = e^{-iS^y \pi/4 } |k' \rangle  $
which is an eigenstate of the $ S^x$ operator
\begin{align}
S^x |k' \rangle^{(x)} = (2k'-N) |k' \rangle^{(x)} .  
\label{sxeigenstate}
\end{align}
The other two spin operators satisfy the commutation relations 
\begin{align}
[S^y, S^z ] = 2i S^x .
\end{align}
Taking the expectation value of $ S_x $ we may then construct position and momentum variables as
\begin{align}
x & \approx \frac{S^z}{\sqrt{2N}} \label{hpapproxx} \\
p &  \approx - \frac{S^y}{\sqrt{2N}}
\label{hpapprox}
\end{align}
where we have assumed an extremal state with $ k' \gg N/2 $ (and $ 0 \le k' \le N$) here. We will consider other cases later.  This definition gives the canonical commutation relations $ [x,p] = i $.  

The number states also are total angular momentum eigenstates satisfying 
\begin{align}
 [(S^x)^2 + (S^y)^2 +(S^z)^2] | k'\rangle^{(x)} = N(N+2) | k'\rangle^{(x)}
\end{align}
which is the Casimir invariant.   Substituting  (\ref{sxeigenstate}) we have 
\begin{align}
[ (S^y)^2 +  (S^z)^2 ] | k'\rangle^{(x)} = 2(N + 2k'N - 2k'^2 ) | k'\rangle^{(x)} 
\label{spinschrodinger}
\end{align}
Using the approximation (\ref{hpapprox}) we may then write
\begin{align}
\frac{1}{2} ( p^2 + x^2 ) | k'\rangle^{(x)} = E_{k'} | k'\rangle^{(x)} 
\end{align}
where ``energy'' is 
\begin{align}
E_{k'} = k' + \frac{1}{2} - \frac{k'^2}{N} 
\label{energyho}
\end{align}
which for $ k' \ll N/2 $  resembles the standard energy eigenstates of the harmonic oscillator with quantum number $ n = k'$.  The expression (\ref{energyho}) is symmetric under reflection $ k' \rightarrow N- k' $ such that $ E_{N-k'} = E_{k'} $.  Therefore for $ k' \gg N/2 $ the energies resemble the expression for the harmonic oscillator with $ n = N - k' $.  

We expect that for $ k' \gg N/2 $,  the eigenstates $ |k' \rangle^{(x)} $ closely resemble harmonic oscillator eigenstates, which are defined as 
\begin{align}
\Phi_n (x) = \langle x | n \rangle = \frac{e^{-x^2/2 } H_n (x) }{2^{n/2} \pi^{1/4}\sqrt{n!}  },
\label{howavefunction}
\end{align}
where $ H_n (x) $ are the Hermite polynomials. Here $ |n \rangle $ are the eigenstates of the harmonic oscillator and $ |x \rangle $ are eigenstates of the position operator $ x $.  Using the mappings $ n = N-k' $ and (\ref{hpapproxx}) we then have 
\begin{align}
 \langle k | k' \rangle^{(x)}  \approx \Phi_{N-k'} \left (\frac{2k-N}{\sqrt{2N}} \right)  .  
 \label{approx1}
\end{align}
whcih is valid for $ k' \gg N/2 $.  

Examples of this approximation are shown in Fig. \ref{fig:app1}(a)(b).  We see that for $ k' \sim N $ the approximation works well, but as $ k' \sim N/2 $, deviations begin to occur.  The reason for this is clear from Fig. \ref{fig:app1}(b), using (\ref{approx1}), the harmonic oscillator wavefunctions begin to exceed the valid range of the $ | k \rangle $ with $ k \in [0,N] $.

The approximations for $ k' \ll N/2$ can be deduced by symmetry of the expression (\ref{rmatrix}).  We have
\begin{align}
    \langle k | N- k' \rangle^{(x)} = (-1)^k \langle k | k' \rangle^{(x)} .
\end{align}
Then repeating the above steps we have for $ k'\ll N/2 $, 
\begin{align}
 \langle k | k' \rangle^{(x)}  \approx (-1)^k \Phi_{k'} \left(\frac{2k-N}{\sqrt{2N}} \right)  .  
 \label{approx2}
\end{align}
where we used the mapping $ n = k' $.  

We can combine (\ref{approx1}) and (\ref{approx2}) into one expression by defining (\ref{nkdef}) which gives our final approximation valid for all $k' $
\begin{align}
 \langle k | k' \rangle^{(x)}  \approx  (-1)^{k \lfloor 1- k'/N \rceil} \left( \frac{2}{\pi} \right)^{1/4} \Phi_{n_{k'}} \left( \frac{2k-N}{\sqrt{2N }} \right)  .  
 \label{finalapprox}
\end{align}
and we have added a normalization factor and $ \lfloor \rceil $ rounds to the nearest integer.

 \begin{figure}%
\includegraphics[width=\linewidth]{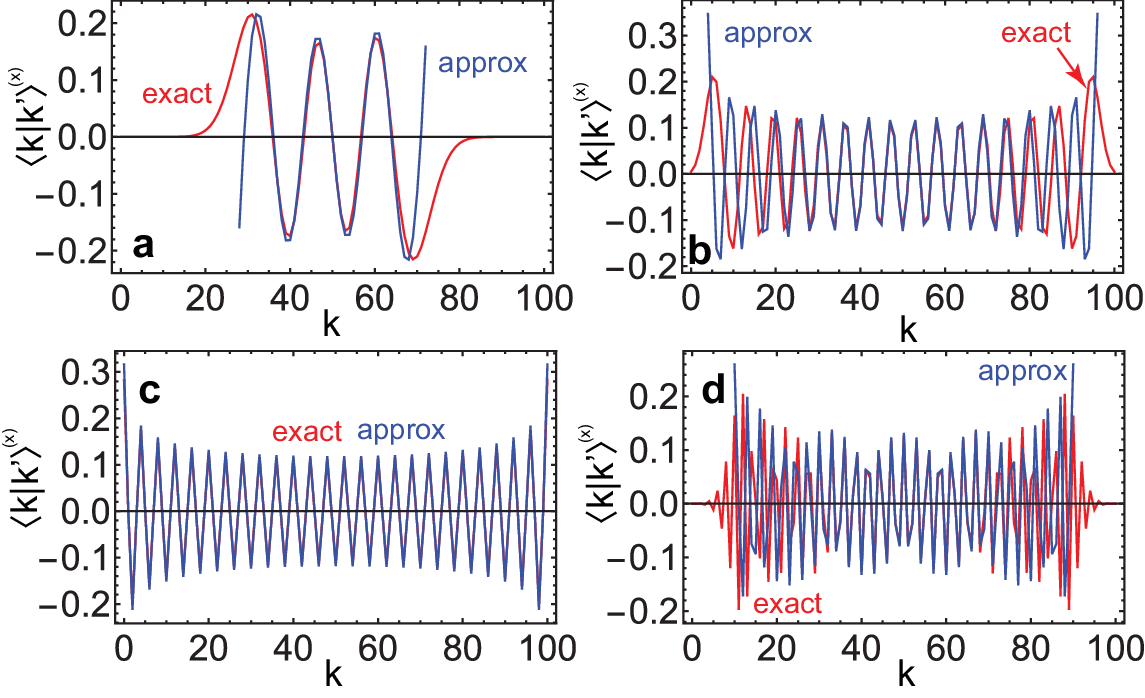}
	\caption{Approximating the inner product $ R_{k',k} = \langle k | k' \rangle^{(x)} $ with the approximate function (\ref{finalapproxOmega}).  Exact curves calculated using (\ref{rmatrix}) are shown for comparison.   Parameters chosen are (a) $ k' = 95 $; (b) $ k' =70 $; (c) $ k' =50 $; (d) $ k' =20 $.   We use $ N  = 100 $ for all plots.  }
	\label{fig:app2}%
\end{figure}

\section{WKB approximation of harmonic oscillator eigenstate}

Using the WKB approximation, the harmonic oscillator eigenstates can be approximated as \cite{grunwald1971wkb}
\begin{align}
\Phi_n (x) \approx \frac{\cos [ \frac{n\pi}{2}  + \frac{x}{2} 
\sqrt{ 2 E_n^{\text{HO}} - x^2 } +E_n^{\text{HO}} \sin^{-1} (\frac{x}{\sqrt{2 E_n^{\text{HO}}}} ) ]}{(2 E_n^{\text{HO}} -x^2)^{1/4}  }
\end{align}
where $ E_n^{\text{HO}} = n + 1/2 $ is the harmonic oscillator energy eigenvalues. This gives virtually perfect  agreement with the exact wavefunctions (\ref{howavefunction}) in the classically allowed region.  

We may obtain a simplified expression for the WKB wavefunction by expanding the argument of the cosine to linear terms in $ x $
\begin{align}
\Phi_n (x) \approx \frac{\cos [ \frac{n\pi}{2}  + \sqrt{ 2 E_n^{\text{HO}} }x 
 ]}{(2 E_n^{\text{HO}} -x^2)^{1/4}  }
 \label{expandedwkb}
\end{align}
The effect of dropping higher order terms in $ x $ is that the oscillations in the harmonic oscillator function become evenly spaced.  For example, in Fig. \ref{fig:app1}(b) notice that the frequency of the oscillations decrease towards the edges.  This introduces deviations in the context of the harmonic oscillator wavefunctions, but relative to to the inner product $ \langle k | k' \rangle^{(x)} $ it is beneficial since the oscillations tend to be more evenly spaced.  

Using (\ref{expandedwkb}) in (\ref{finalapprox}) gives a reasonable approximation to $  \langle k | k' \rangle^{(x)}$ but some empirical adjustments can be made to improve the approximation.  First, the factor of $ \sqrt{ E_n^{\text{HO}}/N } $ can be closely approximated by 
\begin{align}
\sqrt{\frac{n_{k'} +1/2}{N}} \approx \cos^{-1} \left( \left| \frac{2 k'}{N} -1 \right|\right) .
\label{wkbexpression}
\end{align}
We find that the $ \cos^{-1} $ function gives overall better results than the left hand side of the above expression. 

Second, as we have seen in Fig. \ref{fig:app1}(b), the harmonic oscillator wavefunction tends to be broader the expression for $  \langle k | k' \rangle^{(x)}$.  We therefore adjust the denominator in (\ref{wkbexpression}) to account for this.  We perform this by examining the classically allowed regions of the wavefunctions.  Recall that for a harmonic oscillator the classically allowed regions are
\begin{align}
    |x| \le \sqrt{2n + 1 }
\end{align}
Viewing (\ref{spinschrodinger}) as a Schrodinger-like equation, the classically allowed region of the spin $ S^z$ is 
\begin{align}
    |S^z| \le \sqrt{2(N + 2k'N - 2k'^2 )}
\end{align}
We then define the amplitude function
\begin{align}
A_{k',k} = \left( \frac{2(N + 2k'N - 2k'^2 ) - (2k-N)^2}{N} \right)^{-1/4}
\label{amplitudefunc}
\end{align}
Putting together (\ref{wkbexpression}) and (\ref{amplitudefunc}) we obtain our final expression for the approximated inner product
\begin{align}
 \langle k | k' \rangle^{(x)}  =  R_{k', k} \approx
 A_{k',k} \cos [ \frac{n_{k'} \pi}{2}  + \frac{\phi_{k'}}{2}  (2k-N)  ]
 \label{finalapproxOmega}
\end{align}
where $ \phi_{k}$ is defined in (\ref{phidef}).  

Some examples of the approximation are shown in Fig. \ref{fig:app2}.  We see that as with (\ref{finalapprox}) for $ k' \sim 0,N $ the approximation works well.  Remarkably, the approximation gives nearly perfect agreement for  $ k ' \sim N/2$, where (\ref{finalapprox}) tends to fail.  The worst performing regions are near $ k ' \sim N/4, 3N/4 $ where near the edges the there is a frequency deviation which is not accounted for due to the linear approximation (\ref{expandedwkb}).

\section{Decoherence}
\label{app:deco}

We use the formulation following Ref. \cite{pyrkov2014full} to apply the Lindblad dephasing master equation (\ref{lindblad}) using a stochastic rotation, such that starting from an initial state $ \rho = \rho_0 $, evolves as
\begin{align}
\rho(t) = \frac{1}{\sqrt{ 2 \pi \gamma t }} \int_{-\infty}^{\infty} d \xi \exp \left( - \frac{\xi^2}{2 \gamma t} \right) e^{i \xi S^x } \rho_0  e^{ -i \xi S^x }  .
\end{align}
This has the physical interpretation of a mixture of random phase rotations. 

We first apply the dephasing to the initial state $\rho_0 = | \Psi_0 \rangle \langle \Psi_0 | $, where the state $ | \Psi_0 \rangle $ is (\ref{initialstate}). Applying the random phases to all three ensembles, we have
\begin{align}
| \Psi_0 ( \xi_1,\xi_2,\xi_3) \rangle & = e^{i \xi_1 S^x_1 }e^{i \xi_2 S^x_2 } e^{i \xi_3 S^x_3 } | \Psi_0 \rangle \nonumber \\
& = | \psi_0 (\xi_1) \rangle \otimes e^{i (\xi_2 + \xi_3) S^x_3 } | \Psi \rangle \rangle ,
\end{align}
where we defined a rotated version of the initial state $ | \psi_0 (\xi_1) \rangle  = e^{i \xi_1 S^x_1 }  | \psi_0 \rangle $ and use the basis invariant property of the maximally entangled state \cite{kitzinger2020}.  

For Protocol I, we then apply the measurements as given in (\ref{step3protocol1}).  Since the measurements are only applied on ensembles 1 and 2, we may easily write
\begin{align}
& | \Psi_I ( \xi_1,\xi_2,\xi_3)  \rangle  =  \Pi_{k_1}^{x,1} \Pi_{k_2}^{x,2}  P_{\Delta}^{z} | \Psi_0 ( \xi_1,\xi_2,\xi_3) \rangle \nonumber \\
& = e^{i (\xi_2 + \xi_3) S^x_3 }  | \Psi_I ( \xi_1)  \rangle  ,
\label{psi1deco}
\end{align}
where $ \psi_k^{(0)} (\xi_1) = \langle k |  \psi_0 (\xi_1) \rangle $ and 
\begin{align}
&  | \Psi_I ( \xi_1)  \rangle  =   \frac{1 }{\sqrt{N+1}} \nonumber \\
 & \times \sum_{k=\max(0,-\Delta)}^{\min(N,N-\Delta) } 
\psi_k^{(0)} (\xi_1) R_{k_1,k} R_{k_2,k+\Delta}  |k+\Delta \rangle_{3} 
\end{align}
is the state after Step 3 of Protocol I with a rotated initial state.  

We then apply decoherence again to (\ref{psi1deco}) to all three ensembles with the random phases $ \xi_1',\xi_2',\xi_3' $.  At this point ensembles 1 and 2 are disentangled and hence the random phases have no effect, giving the final state
\begin{align}
& | \Psi_I ( \xi_1,\xi_2,\xi_3, \xi_3' )  \rangle  = e^{i (\xi_2 + \xi_3+ \xi_3') S^x_3 } | \Psi_I ( \xi_1)  \rangle   ,
\label{psi2deco}
\end{align}

The final dephased state for Protocol I is 
\begin{align}
\rho_I =  \frac{1}{(2 \pi \gamma t)^2} \int_{-\infty}^{\infty} d \xi_1 d \xi_2 d \xi_3 d \xi_3' e^{- \frac{\xi^2_1 + \xi^2_2 + \xi^2_3  + (\xi_3')^2 }{2 \gamma t} } \nonumber \\
\times | \Psi_I ( \xi_1,\xi_2,\xi_3, \xi_3' )  \rangle \langle \Psi_I ( \xi_1,\xi_2,\xi_3, \xi_3' )  | .
\end{align}

The expectation value of total spin operators can be evaluated as
\begin{align}
& \langle S^z_3 \rangle = \text{Tr} (\rho_I S^z_3 )  \nonumber \\
& =  \frac{1}{(2 \pi \gamma t)^2} \int_{-\infty}^{\infty} d \xi_1 d \xi_2 d \xi_3 d \xi_3' e^{- \frac{\xi^2_1 + \xi^2_2 + \xi^2_3  + (\xi_3')^2 }{2 \gamma t} } \nonumber \\
& \times \langle \Psi_I ( \xi_1)  |   e^{-i (\xi_2 + \xi_3+ \xi_3') S^x_3 }  S^z_3  e^{i (\xi_2 + \xi_3+ \xi_3') S^x_3 } | \Psi_I ( \xi_1)  \rangle \nonumber \\
& =  \frac{1}{(2 \pi \gamma t)^2} \int_{-\infty}^{\infty} d \xi_1 d \xi_2 d \xi_3 d \xi_3' e^{- \frac{\xi^2_1 + \xi^2_2 + \xi^2_3  + (\xi_3')^2 }{2 \gamma t} } \nonumber \\
& \times \langle \Psi_I ( \xi_1)  | [\cos (\xi_2 + \xi_3+ \xi_3') S^z_3 + \sin (\xi_2 + \xi_3+ \xi_3') S^z_3 ] | \Psi_I ( \xi_1)  \rangle \nonumber \\
& = \frac{e^{-3 \gamma t /2}}{\sqrt{2 \pi \gamma t}} \int_{-\infty}^{\infty} d \xi_1 \langle \Psi_I ( \xi_1) | S^z_3 | \Psi_I ( \xi_1)  \rangle ,
\label{lastlined7}
\end{align}
where we used
\begin{align}
\frac{1}{\sqrt{2 \pi \gamma t}} \int_{-\infty}^{\infty} d \xi  \exp \left( - \frac{\xi^2}{2 \gamma t} \right) \cos \xi &  = e^{-\gamma t /2} \\
\frac{1}{\sqrt{2 \pi \gamma t}} \int_{-\infty}^{\infty} d \xi  \exp \left( - \frac{\xi^2}{2 \gamma t} \right) \sin \xi &  =  0
\end{align}
and standard trigonometric identities.  

Finally, in evaluating (\ref{lastlined7}) let us use the fact that according to the results (\ref{prot1error}) we have $ \theta_{\text{tel}} = \theta_0, \phi_{\text{tel}} = \phi_0  $.  Then to a good approximation
\begin{align}
\langle \Psi_I ( \xi_1) | S^z_3 | \Psi_I ( \xi_1)  \rangle & \approx \langle \Psi_I  | e^{-i \xi_1 S^x} S^z_3 e^{-i \xi_1 S^x} | \Psi_I  \rangle \nonumber \\
& = \cos \xi_1 \langle \Psi_I  |  S^z_3 | \Psi_I  \rangle + \sin \xi_1 \langle \Psi_I  |  S^y_3 | \Psi_I  \rangle .
\end{align}
Using this, we may evaluate the integral (\ref{lastlined7}) to obtain
\begin{align}
 \langle S^z_3 \rangle = e^{-2 \gamma t} \langle \Psi_I  | S^z_3 | \Psi_I  \rangle .
\end{align}
The other spin expectations may be similarly evaluated as
\begin{align}
 \langle S^y_3 \rangle  & = e^{-2 \gamma t} \langle \Psi_I  | S^y_3 | \Psi_I  \rangle \nonumber \\
  \langle S^x_3 \rangle &  = \langle \Psi_I  | S^x_3 | \Psi_I  \rangle .
\end{align}

Using the same methods we obtain the similar spin expectation values for the dephased version of Protocol II
\begin{align} 
 \langle S^x_3 \rangle & = \langle \Psi_{II}  | S^x_3 | \Psi_{II}  \rangle \nonumber \\
 \langle S^y_3 \rangle & = e^{-2 \gamma t} \langle \Psi_{II}  | S^y_3 | \Psi_{II}  \rangle \nonumber \\
  \langle S^z_3 \rangle & = e^{-2 \gamma t} \langle \Psi_{II}  | S^z_3 | \Psi_{II}  \rangle .
\end{align}

\bibliography{ref}

\end{document}